\documentclass[%
 reprint,
nofootinbib,
 amsmath,amssymb,
 aps,
prb,
]{revtex4-1}

\usepackage{graphicx}
\usepackage{dcolumn}
\usepackage{bm}
\usepackage{hyperref}

\usepackage[mathlines]{lineno}

\usepackage{color}

\usepackage{xcolor}
\begin{document}
\preprint{arXiv}

\title{
  Efficient One-Loop-Renormalized Vertex Expansions with\\ Connected Determinant Diagrammatic Monte Carlo
}


\author{
  Fedor \v{S}imkovic IV$^{1,2}$}

\author{  Riccardo Rossi$^3$}
\email{riccardorossi4@gmail.com}

\author{  Michel Ferrero$^{1,2}$}
\affiliation{$^1$CPHT, CNRS, Ecole Polytechnique, Institut Polytechnique de Paris, Route de Saclay, 91128 Palaiseau, France\\
  $^2$Coll\`ege de France, 11 place Marcelin Berthelot, 75005 Paris, France\\
 $^3$Center for Computational Quantum Physics, Flatiron Institute, 162 5th Avenue, New York, NY 10010 }

\date{\today}

\begin{abstract}
We present a technique that enables the evaluation of perturbative expansions
based on one-loop-renormalized vertices up to large expansion
orders. Specifically, we show how to compute large-order corrections to the
random phase approximation in either the particle-hole or particle-particle channels.
The algorithm's efficiency is achieved by the summation over contributions of all symmetrized Feynman diagram topologies using determinants, and by integrating out analytically the two-body long-range interactions in order to yield an effective zero-range interaction. Notably, the exponential scaling of the algorithm as a function of perturbation order leads to a polynomial scaling of the approximation error with computational time for a convergent series. To assess the performance of our approach, we apply it to the non-perturbative regime of the square-lattice fermionic Hubbard model away from half-filling and report, as compared to the bare interaction expansion algorithm, significant improvements of the Monte Carlo variance as well as the convergence properties of the resulting perturbative series.
\end{abstract}

\maketitle

\section{Introduction}
\label{sec_intro}

In recent years, there has been a growing need for controllable numerical
techniques in the field of strongly
correlated systems in order to reliably predict the collective behavior of
electrons in solids and establish a connection with experiments~\cite{leblanc2015solutions, schafer2020tracking}. Simultaneously, multiple novel experimental realizations of strongly correlated models by means of cold atoms on optical lattices have not only provided a way of testing numerical approaches on a qualitative level, but have also increased the importance of producing quantitatively accurate results~\cite{jaksch1998cold, Bloch_review_2005,
kohl2005fermionic,lewenstein2007ultracold, jordens2008mott,
schneider2008metallic, hulet2015antiferromagnetism, greif2015formation,
parsons2016site, cheuk2016observation, greiner2017, nichols2019spin, hartke2020measuring}.

The Diagrammatic Monte Carlo approach~\cite{ProkSvistFrohlichPolaron,
  ProkofevSvistunovPolaronLong, van2010diagrammatic, kris_felix, deng, vsimkovic2019superfluid}
is a method that has recently made progress in this regard.
It is based on the stochastic sampling of Feynman diagrams
directly in the thermodynamic (and possibly continuum) limit and is numerically-exact
when extrapolation to infinite diagram order is possible. In its original
formulation, the method uses a Monte Carlo algorithm to compute
contributions from individual Feynman diagram topologies.
Despite many recent advancements~\cite{wu_controlling, kun_chen, gull_inchworm,
vucicevic2019real, taheridehkordi2019algorithmic, taheridehkordi2019optimal},
this approach fundamentally suffers from large variance induced by
the almost-exact cancellation of a factorially-increasing number of diagrams
 as a function of expansion order.

At thermal equilibrium, this issue has been overcome by the development of the
Connected Determinant Diagrammatic Monte Carlo algorithm (CDet)~\cite{cdet} and
its one-particle irreducible extensions~\cite{fedor_sigma, alice_michel,
rr_sigma} which at each Monte Carlo step sum the full factorial number of possible bare connected, or irreducible, diagram topologies in the spacetime representation at only exponential computational cost. This has been shown to lead to a polynomial scaling of the
error bar with respect to the computational time for
observables within the convergence radius of the perturbative series~\cite{rr_epl}.
Thanks to these improvements in computational complexity, unprecedentedly high expansion
orders have been reached ($\gtrsim\!10$), allowing for the evaluation of series well
beyond their radius of convergence~\cite{fedor_sigma}. Similarly effective
exponential algorithms overcoming the factorial barrier have also been found
for the real-time evolution of quantum systems~\cite{olivier, corentin,
kid_gull_cohen, moutenet2019cancellation, mavcek2020quantum}.

Diagrammatic Monte Carlo using the bare interaction expansion has allowed for
important progress in the study of fermionic systems on a lattice at finite
temperature~\cite{kozik2010diagrammatic, wu_controlling, fedor_sigma,
fedor_hf, kim_cdet, lenihan2020entropy}, but it still has its limitations.  For
example, it has been documented that poles, which can severely limit the radius of convergence, can appear in the complex plane of the evaluated functions. Specifically, such poles have been shown to appear in the two-dimensional Fermi-Hubbard model: On the negative real axis of the complex plane where they are related to a superfluid phase transition in the attractive Fermi-Hubbard model~\cite{cdet} as well as in the vicinity of the positive real axis and related to sharp crossovers due to the onset of strong magnetic fluctuations~\cite{fedor_sigma, fedor_hf}. Further, at very low temperatures,
infrared divergencies are expected to appear~\cite{feldman}. Another
limitation of the bare interaction series, as the temperature is lowered, is the rapid increase in Monte Carlo variance which is due to wider spatial spread of interaction vertices in the spacetime representation of Feynman diagrams.

It is, therefore, evident that further progress in this approach must come
from evaluating more advanced perturbative expansions in order to improve the analytic properties of perturbative series. It has been shown that the renormalization of the chemical potential can already lead to substantial improvements of the
complex plane structure for evaluated series~\cite{rubtsov2005continuous, olivier, wu_controlling}.
In Ref.~\onlinecite{rossi2020renormalized} a general renormalization technique was introduced within the determinantal formalism, and it has been shown therein that one-particle renormalization is essential to reach deep into the pseudogap regime of the doped two-dimensional Hubbard model.

In this work we illustrate how one can efficiently go beyong single-particle renormalization within the
determinantal formalism
by systematically computing the corrections to the random phase
approximation (RPA)~\cite{bohm1951collectiveI, pines1952collectiveII, bohm1953collectiveIII}, either in the particle-particle or particle-hole channel. Whilst it is in principle possible to use the general formalism of Ref.~\onlinecite{rossi2020renormalized} for the specific case of vertex renormalization performed in this work, the method we introduce here is more efficient as well as easier to implement, albeit less general. Importantly, the algorithm we present also overcomes the factorial barrier and does not suffer from misleading convergence issues~\cite{shifted_action, kozik2015nonexistence}. We further show that the resulting series can have a larger convergence radius with
respect to the bare interaction series, as we document in the two-dimensional
hole-doped Hubbard model.
The removal of Feynman diagram topologies with RPA bubble insertions from the series also leads to higher locality of vertices in real space, and thus to an improved Monte Carlo variance allowing the algorithm to reach larger expansion orders as compared to the bare interaction series algorithm, despite having higher computational cost. Let us also emphasize that, when considering systems directly in continuous space, performing vertex renormalization is usually an unavoidable step in the process of defining the theory.

The paper is structured as follows: In Sec.~\ref{sec_def} we introduce the notations used in this work. Sec.~\ref{sec_G0P0} provides an in-detail derivation of the theory for the bare RPA particle-particle expansion in both algebraic (Sec.~\ref{subsec_G0P0_shifted_action}, \ref{subsec_G0P0_HS}, \ref{subsec_G0P0_fermionic_action}) and diagrammatic formulations (Sec.~\ref{subsec_G0P0_feynman}). In Sec.~\ref{sec_G0P0_cdet} we introduce the determinantal algorithm which allows the computation of the expansion up to large orders, discussing in particular the analytical integration over the long-range part of the interaction vertices (Sec.~\ref{subsec_integration}). We also briefly describe the few technical modifications needed to perform an RPA expansion without Hartree insertions in Sec.~\ref{sec_G1P1}. Finally, we present benchmark numerical results obtained for the Hubbard model in Sec.~\ref{sec_results} in the single-site model as well as on the two-dimensional square-lattice (Sec.~\ref{subsec_atom} and \ref{subsec_2d_hubbard}, respectively).

\section{Definitions and notations}
\label{sec_def}

\subsection{Hubbard model Hamiltonian}
\label{subsec_hubbard}

In what follows, we focus on the two-dimensional fermionic Hubbard model~\cite{hubbard1963electron,anderson1963theory,anderson1997theory}, defined by the grand-canonical Hamiltonian
\begin{equation}\label{eq_hamiltonian}
\hat{H} := \sum_{\mathbf{k},\sigma} \left(\epsilon_{\mathbf{k}} -\mu_\sigma \right)c_{\mathbf{k}\sigma}^\dagger c_{\mathbf{k}\sigma}^{\phantom{\dagger}}+U\sum_{\mathbf{r}} n_{\mathbf{r}\uparrow}n_{\mathbf{r}\downarrow},
\end{equation}
where $c_{\mathbf{k}\sigma}^\dagger$ ($c_{\mathbf{k}\sigma}^{\phantom{\dagger}}$) creates (annihilates) a fermion of spin $\sigma \in \{ \uparrow, \downarrow \} $ and momentum $\mathbf{k}$, $\mu_{\sigma}$ denotes the chemical potential, $U$ the onsite repulsion strength, $\mathbf{r}$ labels lattice sites, and the (square lattice) dispersion is given by
\begin{align}
\epsilon_{\mathbf{k}}=-2t\left(\cos k_x+\cos k_y\right)-4t^{\prime}\cos k_x\cos k_y,
\end{align}
where $t$ and $t^{\prime}$ are the nearest-neigbor and next-nearest-neighbor hopping amplitudes, respectively. In the following, we measure quantities in units of $t$ by taking $t=1$.

\subsection{Action representation}
\label{subsec_action}

We consider the action formulation of the Hamiltonian~\eqref{eq_hamiltonian} in the imaginary time representation:
\begin{equation}\label{eq_action_hubbard}
S_{\text{phys}}=S^{\text{F}}_{0}+S_{I},
\end{equation}
where the non-interacting fermionic term of the action is given by
\begin{equation}\label{eq_action_F_0}
  S_{0}^{F}=-\sum_{\sigma}
\int_X
  \bar{\psi}_\sigma(X)
  \left((G_{0}^\sigma)^{-1}\psi_\sigma\right)(X),
\end{equation}
the interaction term is given by
\begin{equation}\label{eq_action_I}
  S_{I}=U\,\int_X\left(\bar{\psi}_\uparrow \bar{\psi}_\downarrow \psi_\downarrow \psi_\uparrow\right)(X),
\end{equation}
and the non-interacting (bare) Green's function is
\begin{equation}\label{eq_G0}
  G_{0}^\sigma(K)=
  \frac{1}{i\omega_m-\epsilon_{\mathbf{k}}+\mu_{\sigma}},
\end{equation}
where $X:=(\mathbf{r},\tau)$ is a spacetime coordinate, $\tau\in[0,\beta]$ is the imaginary time where $\beta$ is the inverse temperature, $\psi_\sigma(X)$ is a Grassman-variable valued spacetime field, $K:=(\mathbf{k},i\omega_m)$ is the momentum-frequency, $\omega_m:=(2m+1)\pi/\beta$, $m\in\mathbb{Z}$, is a fermionic Matsubara frequency, and the integral over spacetime variables means sum over lattice sites and integration over imaginary time
\begin{equation}\label{}
\int_X :=\sum_{\mathbf{r}}\int_0^\beta d\tau\,.
\end{equation}

\subsection{Connected Determinant Monte Carlo for the bare interaction expansion}
\label{subsec_cdet}

Before we describe the vertex renormalization, we first give a brief
recapitulation of the individual steps of the CDet algorithm~\cite{cdet} for the bare interaction expansion.
For
simplicity of presentation, we focus our discussion on the calculation of the perturbative series
of the grand-canonical potential density $\Phi_G$ (equal to minus the pressure for a homogeneous system):
\begin{equation}
  \Phi_G:= -\frac{\log \,\text{Tr}\; e^{-\beta\hat{H}}}{\beta L_x L_y} =
  \Phi_{G}(U=0)+\sum_{n=1}^\infty U^n\,\phi_n^{\text{bare}},
\label{grand_potential_series}
\end{equation}
where $L_x$ and $L_y$ are the linear lattice sizes and $\phi_n^{\text{bare}}$ is the sum of all connected diagrams with $n$ internal (bare) $U$ interaction vertices and no external vertices.
The coefficients $\phi_n^{\text{bare}}$ are computed from the stochastic sampling of internal vertices parametrized by $X_j = (\mathbf{r}_j, \tau_j)$, where
$\mathbf{r}_j$ labels a lattice site and $\tau_j\in[0,\beta]$ an imaginary time:
\begin{equation}\label{eq_phi_n}
  \phi_n^{\text{bare}}=\frac{1}{\beta \,L_x\,L_y\,n!}
  \int_{X_1,\dots,X_n}\,
  \,c(\{X_1,\dots,X_n\}),
\end{equation}
where $c(\{X_1,\dots,X_n\})$ is the sum of all connected Feynman diagrams that can be constructed from a set $V$ of bare interaction vertices at spacetime positions $\{X_1,\dots,X_n\} =: V$, symmetrized with respect to the exchange of $X_1,\dots,X_n$. We remark that the spacetime volume factor $\beta\, L_x\, L_y$ in Eq.~\eqref{eq_phi_n} is cancelled by the translation invariance of the integrand.

In order to compute the integral of Eq.~\eqref{eq_phi_n}, one needs to evaluate $c(V)$. To achieve this, one introduces $a(V)$, the sum of all connected and disconnected bare Feynman diagrams that can be built from the vertices in $V$, which, by the Wick's theorem, is given by
\begin{equation}\label{eq_a_M}
  a(V) = (-1)^{n+1}\;\operatorname{det} \left(M_{\uparrow}(V)\right) \operatorname{det} \left(M_{\downarrow}(V)\right),
\end{equation}
where the elements of the $n\times n$ matrices $M_\sigma(V)$ are the bare propagators $G_{0}^{\sigma}$ defined in Eq.~\eqref{eq_G0}
\begin{equation}\label{eq_M}
  \left(M_{\sigma}(V)\right)_{jk} = G_{0}^\sigma(X_j,X_k) = G_{0}^\sigma(X_j - X_k).
\end{equation}
To obtain the sum of all connected diagrams $c(V)$, one needs to
eliminate all disconnected diagrams from $a(V)$ by making use of the recursive
formula:
\begin{equation}\label{eq_recursive_formula}
  c(V) = a(V) - \sum_{\substack{V^\prime \subsetneq V \\ V^\prime \ni X_1}} c(V^\prime) \, a(V \setminus V^\prime),
\end{equation}
where, in order to properly define connectivity, the sum is over all subsets $V^\prime$ containing the arbitrarily chosen vertex $X_1$
from $V$. The integration in Eq~\eqref{eq_phi_n} is then numerically performed
with a Markov-chain Monte Carlo algorithm.

\section{$\mathbf{G_0 P_0^{\text{pp}}}$ expansion}
\label{sec_G0P0}

In this section we solely discuss the RPA expansion
in the particle-particle channel, as the particle-hole case
can be derived analogously. We use the shifted-action expansion formalism,
introduced in Ref.~\onlinecite{shifted_action}, in order to precisely define the
counterterm action. We then present
the Feynman-diagrammatic rules for this expansion.

\subsection{Shifted-action expansion formalism}
\label{subsec_G0P0_shifted_action}

In this section we briefly present the shifted-action formalism
introduced in Ref.~\onlinecite{shifted_action}.
We start from the action of the Hubbard model,
defined by Eq.~\eqref{eq_action_hubbard},
and we introduce a Hubbard-Stratonovich
bosonic field $\eta$ coupled to $(\psi_\downarrow \psi_\uparrow)(X)$.
We can then rewrite the interaction part of the action~\eqref{eq_action_I} as
\begin{equation}\label{eq_action_HS}
S_{\text{phys}}^{\text{HS}}=S^{\text{F}}_{0}+S^{\text{HS}}_{0}+S^{\text{HS}}_{I},
\end{equation}
where
\begin{equation}\label{eq_action_HS_0}
S^{\text{HS}}_{0}:=\frac{1}{U}\int_X (\bar{\eta}\eta)(X),
\end{equation}
\begin{equation}\label{eq_action_HS_I}
S^{\text{HS}}_{I}:=i\int_X \bar{\eta}(X)\;(\psi_\downarrow \psi_\uparrow)(X)+h.c.
\end{equation}
where $i$ is the imaginary unit.
In order to define the diagrammatic expansion, it is useful to introduce
a formal expansion parameter $\xi$, such that the expansion in $\xi$
reproduces order by order the diagrammatic expansion.
We introduce therefore a $\xi$-dependent action $S^{\text{HS}}(\xi)$:
\begin{equation}\label{eq_action_HS_xi}
S^{\text{HS}}(\xi) = S^{\text{F}}_{0}(\xi)+S^{\text{HS}}_{0}(\xi)+S^{\text{HS}}_{I}(\xi).
\end{equation}
Every quantity, such as the grand-canonical potential density $\Phi_G$, for instance, can be expanded in powers of $\xi$:
\begin{equation}\label{eq_phi_xi}
\Phi_G(\xi) =\Phi_G(\xi=0)+\sum_{n=1}^\infty \xi^n\,\phi_n^{\text{expansion}},
\end{equation}
where $\phi_n^{\text{expansion}}$ is the contribution of all
order $n$ Feynman diagrams of a particular diagrammatic expansion.
We further impose that for $\xi=1$ one gets back the physical action, defined in Eq.~\eqref{eq_action_HS}:
\begin{equation}
S^{\text{HS}}(\xi=1)=S_{\text{phys}}^{\text{HS}}.
\end{equation}

\subsection{Hubbard-Stratonovich shifted action}
\label{subsec_G0P0_HS}

In this section, we give explicit expressions
for $S^{\text{HS}}(\xi)$ for the particle-particle RPA bare expansion,
which we denote the $G_0\,P_0^{\text{pp}}$ expansion.
We consider the Hartree shift of the chemical potential,
which consists of adding a linear in $\xi$
term to the non-interacting action~\eqref{eq_action_F_0} which is
proportional to the particle number:
\begin{equation}\label{eq_action_F_0_xi}
  \begin{split}
S^{\text{F}}_{0}(\xi)&:=  -\sum_{\sigma}
\int_X \bar{\psi}_\sigma(X) \left((G_{0}^\sigma)^{-1}\psi_\sigma\right)(X)\\
&-\xi\, U\,\sum_{\sigma}n_0^{\bar{\sigma}}\;\int_X (\bar{\psi}_{\sigma} \psi_\sigma)(X),
  \end{split}
\end{equation}
where $X$ is a spacetime coordinate, $\sigma\in\{\uparrow,\downarrow\}$ is the spin, $\bar{\sigma}$ is the opposite spin to $\sigma$,
and $n_0^{\sigma}:=G_0^\sigma(\mathbf{r}=0,\tau=0^{-})$
is the non-interacting density.

We now give the expression for the quadratic-in-$\eta$ part of the action
in the particle-particle ladder renormalization expansion:
we introduce a $\xi$-dependent term to the action~\eqref{eq_action_HS_0}
that cancels the first contribution to the pair-self energy of the field $\eta$
\begin{equation}\label{eq_action_HS_0_xi}
  \begin{split}
    &S^{\text{HS}}_{0}(\xi):= \frac{1}{U}\int_X (\bar{\eta}\eta)(X)+\\
    &+(1-\xi)\int_{Y,X} \bar{\eta}(Y)\,(G_0^\uparrow G_0^\downarrow)(Y,X)\,\eta(X),
    \end{split}
\end{equation}
where $X$ and $Y$ are spacetime coordinates.
We see that for $\xi=0$, $S^{\text{HS}}_{0}(\xi)$ contains the inverse of the RPA particle-particle propagator, and that the linear term in $\xi$ is a counterterm that cancels bubble insertions.

The coupling part of the Hubbard-Stratonovich shifted action,
$S^{\text{HS}}_{I}(\xi)$, is
obtained by multiplying the action term ~\eqref{eq_action_HS_I} by $\sqrt{\xi}$:
\begin{align}\label{eq_action_R}
  S^{\text{HS}}_{I}(\xi) = i\sqrt{\xi}\int_X \bar{\eta}(X)\;(\psi_\downarrow \psi_\uparrow)(X)+h.c.
\end{align}
This means that two insertions of $S^{\text{HS}}_{I}$ are necessary in order to generate one vertex.
Using Eqs.~\eqref{eq_action_HS} and \eqref{eq_action_HS_xi}, we see that
\begin{equation}\label{}
S^{\text{HS}}(\xi=1)=S_{\text{phys}}^{\text{HS}}
\end{equation}
which implies that when we evaluate the series of Eq.~\eqref{eq_phi_xi} for $\xi=1$,
we obtain the exact physical result.

\begin{figure}
  \includegraphics[width=0.45\textwidth]{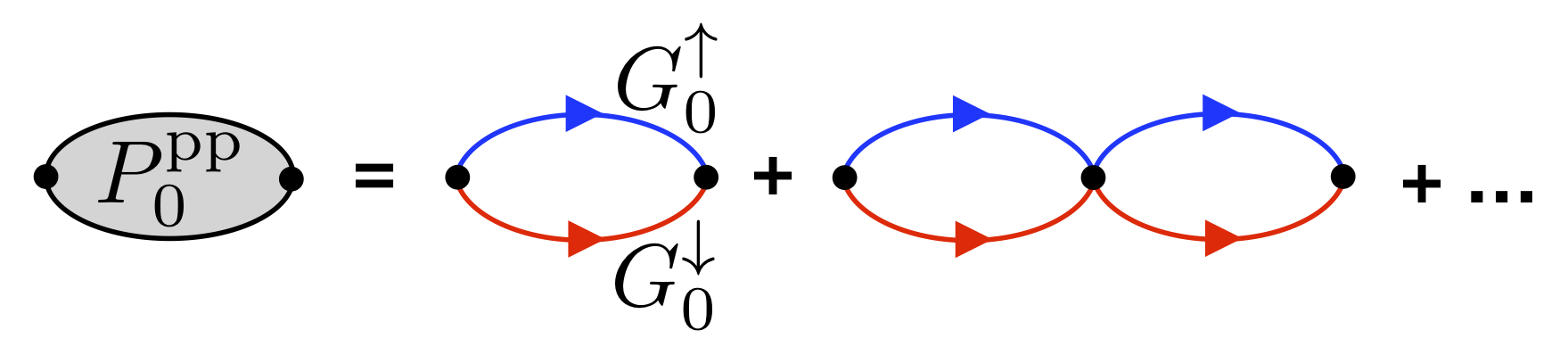}
\caption{Feynman-diagrammatic definition of $P_0^{\text{pp}}$.\label{fig_P_0}}
\end{figure}

\begin{figure}
  \includegraphics[width=0.49\textwidth]{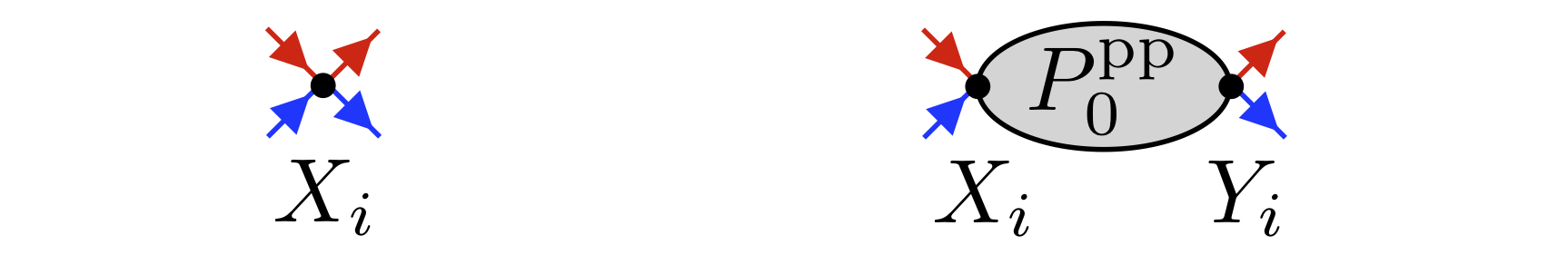}
  \caption{Local vertex $U$ and non-local vertex $P^{\text{pp}}_{0}$ in the RPA particle-particle
    diagrammatic expansion. \label{fig_vertices}}
\end{figure}

\subsection{Fermionic shifted action}
\label{subsec_G0P0_fermionic_action}

We now proceed to integrate out the Hubbard-Stratonovich field $\eta$ in order to obtain a purely
fermionic action. We rewrite Eq.~\eqref{eq_action_HS_0_xi} as
\begin{equation}\label{}
  \begin{split}
  S^{\text{HS}}_{0}(\xi)&=: \int_{Y,X} \bar{\eta}(Y)\,(\Gamma_0^{-1})(Y,X)\,\eta(X)\\
  &-\xi\int_{Y,X} \bar{\eta}(Y)\,(G_0^\uparrow G_0^\downarrow)(Y,X)\,\eta(X),
  \end{split}
\end{equation}
where $\Gamma_0$ is the RPA interaction vertex:
\begin{equation}\label{}
\Gamma_0(X,Y) := U\,\delta(X-Y) + P_0^{\text{pp}}(X,Y),
\end{equation}
and where $P_0^{\text{pp}}(X, Y)$ is the sum of all ladder diagrams and the integrals are over space-time variables $X$ and $Y$.
The graphical definition of $P_0^{\text{pp}}(X, Y) = P_0^{\text{pp}}(X-Y)$ as an infinite series of diagrams is shown in Fig.~\ref{fig_P_0}. We denote the Fourier transform of $P_0^{\text{pp}}(X, Y)$ as $P_0^{\text{pp}}(K)$, which satisfies the following relation:
\begin{equation}
  P^{\text{pp}}_0(K) = U \sum_{n=1}^{\infty} \left( U \tilde{P}_0^{\text{pp}}(K) \right)^{n}
  = \frac{U^2 \tilde{P}_0^{\text{pp}}(K)}{1 - U \tilde{P}_0^{\text{pp}}(K)},
\end{equation}
where
\begin{align}
  \tilde{P}_0^{\text{pp}}(X, Y) := - G^{\uparrow}_0(X,Y) G^{\downarrow}_0(X,Y).
\end{align}

After integrating out the Hubbard-Stratonovich field $\eta$,
we obtain the purely fermionic action $S^{\text{F}}(\xi)$
\begin{equation}\label{eq_action_F_xi}
  S^{\text{F}}(\xi)= S^{\text{F}}_{0}(\xi) +
  S_{\text{I}}(\xi) +
  S_{\text{I}}^{\text{ct}}(\xi),
\end{equation}
where $S^{\text{F}}_{0}(\xi)$ is given by Eq.~\eqref{eq_action_F_0_xi}, the interaction term is
\begin{equation}\label{eq_action_I_xi}
\begin{split}
  S_{I}(\xi)&=\xi\,\int_{Y,X} (\bar{\psi}_{\uparrow} \bar{\psi}_{\downarrow}) (Y)\; \Gamma_0(Y,X)
  \;(\psi_{\downarrow} \psi_{\uparrow}) (X)\\
  &=\xi\,U\int_{X} (\bar{\psi}_{\uparrow} \bar{\psi}_{\downarrow} \psi_{\downarrow} \psi_{\uparrow}) (X) +\\
  &+\xi\,\int_{Y,X} (\bar{\psi}_{\uparrow} \bar{\psi}_{\downarrow}) (Y)\; P_0^{\text{pp}}(Y,X)
  \;(\psi_{\downarrow} \psi_{\uparrow}) (X),
\end{split}
\end{equation}
and the corresponding counterterms in the interaction part of the action become
\begin{equation}\label{eq_action_ct}
  \begin{split}
    S_{I}^{\text{ct}}(\xi)&:=-\sum_{l=1}^\infty (-\xi)^{l+1}\times\\
    &\times\int_{Y,X} (\bar{\psi}_{\uparrow}\bar{\psi}_{\downarrow}) (Y)\, \tilde{P}_{0;l}^{\text{pp}}(Y,X)
\,(\psi_{\downarrow} \psi_{\uparrow}) (X),
\end{split}
\end{equation}
where
\begin{equation} \tilde{P}_{0;l}^{\text{pp}}(K):=\left(\Gamma_0(K)\,\tilde{P}_{0}^{\text{pp}}(K)\right)^l\Gamma_0(K)
\end{equation}
is the $l$-bubble counterterm. It is now easy to verify that when the action $S^{\text{F}}(\xi)$ is evaluated for $\xi =1$, one gets back the physical action $S_\text{phys}$, as defined in Eq.~\eqref{eq_action_hubbard}
\begin{equation}\label{}
  S^{\text{F}}(\xi=1)=S_{\text{phys}}.
\end{equation}

\subsection{Feynman-diagrammatic interpretation}
\label{subsec_G0P0_feynman}

\begin{figure}
  \includegraphics[width=0.49\textwidth]{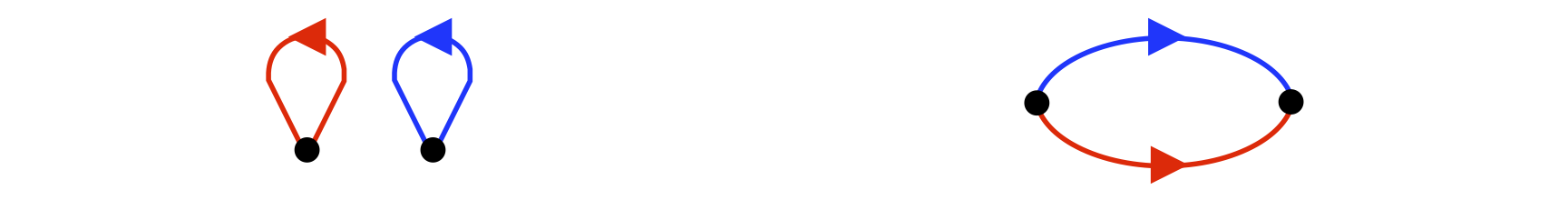}
  \caption{Feynman-diagram insertions that cannot appear in the particle-particle $G_0 P^{\text{pp}}_0$ expansion. This includes local tadpoles (on the left) and the particle-particle bubble (on the right).
    \label{fig_G0P0_removed}}
\end{figure}

\begin{figure}
  \includegraphics[width=0.49\textwidth]{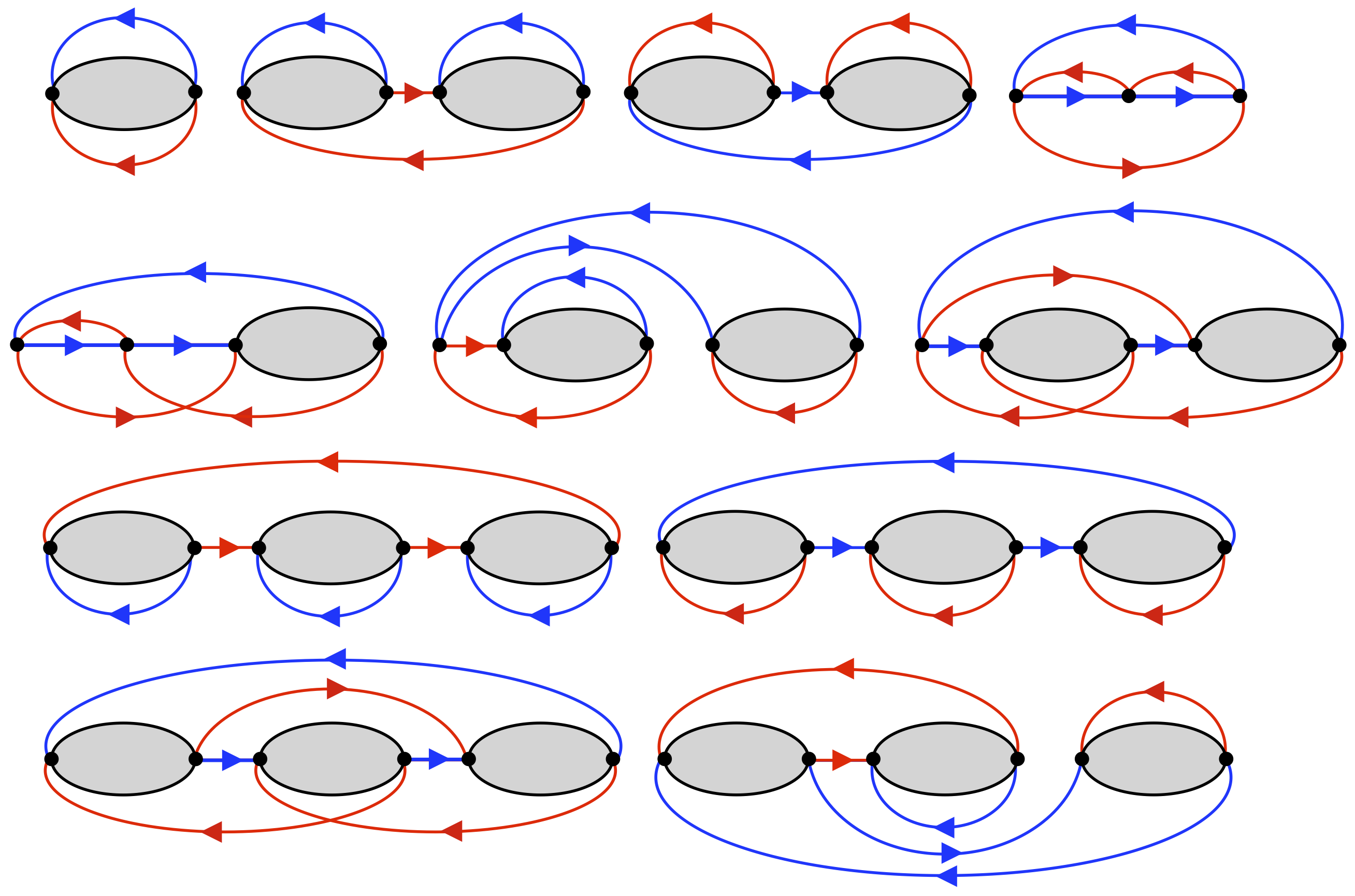}
  \caption{All Feynman diagrams of the $G_{0} P^{\text{pp}}_{0}$ expansion up to 3rd expansion order.
    \label{fig_G0P0_diagrams}}
\end{figure}

We now present the Feynman diagrammatic rules for the $G_0\,P_0^{\text{pp}}$ expansion, defined by
the action~\eqref{eq_action_F_xi}. Equation~\eqref{eq_action_I_xi} defines two types of interaction vertices: a local Hubbard interaction vertex, and a non-local interaction $P_0^{\text{pp}}$ corresponding to the second term. The Feynman diagram definition of these two vertices is presented in Fig.~\ref{fig_vertices}.
In Fig.~\ref{fig_G0P0_removed} we show the diagram insertions that are absent from the expansion: the Hartree shift of the chemical potential, introduced in Eq.~\eqref{eq_action_F_0_xi}, removes tadpole diagrams; the counterterm part of the action, Eq.~\eqref{eq_action_ct}, eliminates all Feynman diagrams with particle-particle bubble insertions. In Fig.~\ref{fig_G0P0_diagrams} we give all Feynman diagrams for the grand-canonical potential density $\Phi_G$ up to third order in $\xi$ for the $G_0 P_0^{\text{pp}}$ expansion.

\section{Connected Determinant Diagrammatic Monte Carlo for the $\mathbf{G_0 P_0^{\text{pp}}}$ expansion}
\label{sec_G0P0_cdet}

In this section we describe how to efficiently perform the $G_0\,P_0^{\text{pp}}$ expansion within the framework of CDet. We start by presenting the expressions obtained by considering $S_I(\xi)$, defined in Eq.~\ref{eq_action_I_xi}, while neglecting the counterterm action $S_{I}^{\text{ct}}(\xi)$, defined in Eq.~\eqref{eq_action_ct}, and the Hartree shift, defined in~\eqref{eq_action_F_0_xi}. We then show how to correct these expressions to take counterterms into account, and finally how to integrate out the non-local interaction vertex.

\subsection{Expansion without counterterms}
\label{subsec_expansion}

Let us consider the order $n$ expansion in $\xi$ of the action term $S_I(\xi)$. For each $\xi$, we can choose either the local vertex, which has one spacetime coordinate $X_j$, or the non-local vertex, which has two spacetime coordinates $X_j$ and $Y_j$, see Eq.~\eqref{eq_action_I_xi} and Fig.~\ref{fig_vertices}. There are $2^n$ such choices. Without loss of generality, we suppose that the first $u$ vertices are local, and the others are non-local:
\begin{equation}\label{}
  W:=\{X_1, \dots, X_u, (X_{u+1},Y_{u+1}),\dots,(X_{n},Y_n)\},
\end{equation}
where $W$ is defined as the set of spacetime positions of the interaction vertices.
  Eq.~\eqref{eq_M} must be modified to take into account the non-locality of some of the vertices:
\begin{equation}\label{eq_M2}
  \left(M_{\sigma}(W)\right)_{jk} = G^{\sigma}_0(X_j,Z_k) = G^{\sigma}_0(X_j-Z_k),
\end{equation}
where
\begin{equation}\label{}
Z_j:=\left\{\begin{array}{ll}X_j\qquad\qquad& \text{for}\;\;j\le u\\ Y_j\qquad&\text{otherwise}\end{array}\right.
\end{equation}
For this particular choice of local and non-local vertices, discarding the counterterms and using Eq.~\eqref{eq_a_M} and~\eqref{eq_recursive_formula}, we get a contribution to the grand-canonical potential density  $\Phi_G$ equal to
\begin{equation}\label{eq_integral_G0P0}
  \frac{\xi^n\,U^u}{\beta L_x L_y n!}\int_{X_1,\dots,X_n,Y_{u+1},\dots,Y_n} c(W)\,\prod_{j=u+1}^n P_0^{\text{pp}}(Y_j-X_j).
\end{equation}

\subsection{Elimination of bare tadpoles}
\label{subsec_tadpoles}

The Hartree shift of the chemical potential, defined by the $\xi$ term in Eq.~\eqref{eq_action_F_0_xi}, is diagrammatically equivalent to eliminating bare tadpoles (Fig.~\ref{fig_G0P0_removed}). We move the Hartree shift term of Eq.~\eqref{eq_action_F_0_xi} to the interaction part of the action, Eq.~\ref{eq_action_I_xi}, to obtain:
\begin{equation}\label{eq_action_I_xi_tilde}
  \begin{split}
  &S_{\tilde{I}}(\xi)=- \xi\, U\,n_0^\uparrow\, n_0^\downarrow\int_X 1\\
    &+\xi\,U\int_{X} ((\bar{\psi}_\uparrow\psi_\uparrow)(X)-n_0^\uparrow)
    ((\bar{\psi}_\downarrow \psi_\downarrow)(X)-n_0^\downarrow) \\
    &+\xi\,\int_{Y,X} (\bar{\psi}_{\uparrow} \bar{\psi}_{\downarrow}) (Y)\; P_0^{\text{pp}}(Y,X)
    \;(\psi_{\downarrow} \psi_{\uparrow}) (X).
\end{split}
\end{equation}
The first term on the r.h.s of Eq.~\eqref{eq_action_I_xi_tilde} is a constant and it can be dropped in most cases; however, for the grand-canonical potential density $\Phi_G$, it contributes at first order.

From a determinantal point of view, it is well known that the chemical potential shift introduced in Eq.~\eqref{eq_action_I_xi_tilde} can be easily taken into account by setting the diagonal of the matrices $M_{\sigma}(W)$ to zero when the entry corresponds to a local vertex~\cite{rubtsov2005continuous}:
 \begin{equation}\label{eq_def_tilde_M}
   (\tilde{M}_\sigma(W))_{jk}:=
   \left\{\begin{array}{ll}(1-\delta_{jk})\;
  (M_\sigma(W))_{jk}\qquad\qquad&\text{for}\;j\le u\\
  (M_\sigma(W))_{jk}&\text{otherwise}\end{array}\right.
\end{equation}

\subsection{Elimination of particle-particle bubbles and generation of unphysical diagrams}
\label{subsec_bubbles}

We discuss here the elimination of diagrams with ladder particle-particle insertions as dictacted by Eq.~\eqref{eq_action_ct}.
In order to do so, it is useful
to observe that the two matrices $\tilde{M}_{\uparrow}$ and $\tilde{M}_{\downarrow}$ can be
transposed and multiplied before the determinant is taken:
\begin{equation}
  \det\left(\tilde{M}_{\uparrow}(W)\right)\, \operatorname{det}\left(\tilde{M}_{\downarrow}(W)\right)
   = \operatorname{det} \left(\tilde{M}_{\uparrow}(W)\,\tilde{M}^{T}_{\downarrow}(W)\right),
\end{equation}
where the new matrix entries are sums of pairs of connected bare Green functions of opposite spin:
\begin{equation}\label{eq_MupMdn_jk}
  \begin{split}
 &\left( \tilde{M}_{\uparrow}(W)\,\tilde{M}^{T}_{\downarrow}(W)\right)_{jk} =\\
    &=  \sum_{l\in\{1,\dots,u\}\setminus\{j,k\}} G^{\uparrow}_0(X_j,X_l)\; G^{\downarrow}_0(X_k,X_l)\\
    &+\sum_{l\in\{u+1,\dots,n\}} G^{\uparrow}_0(X_j,Y_l)\; G^{\downarrow}_0(X_k,Y_l).
  \end{split}
\end{equation}
For some observables, such as the density, the sizes of the matrices $\tilde{M}_{\uparrow}(W)$ and $\tilde{M}_{\downarrow}(W)$ may not be identical. It is then first necessary to pad the smaller matrix by the appropriate number of rows and columns with diagonal entries equal to one and off-diagonal entries equal to zero. This ensures that the correct diagram topologies are generated by the determinant.

As one can see from Eq.~\eqref{eq_MupMdn_jk}, the ladder diagrams participating in the renormalized vertex $P_0^{\text{PP}}$ are all generated by the diagonal of the matrix $\tilde{M}_\uparrow(W)\,\tilde{M}^{T}_\downarrow(W) $.
It would therefore seem natural to simply remove all diagonal entries from the matrix. It turns out, however, that this by itself does not lead to the correct sum of diagram
topologies \footnote{We are grateful to T.~Ohgoe and F.~Werner for pointing out this fact to us.}. Indeed, let us explicitly consider the determinant of the matrix $\tilde{M}_\uparrow(W)\,\tilde{M}^{T}_\downarrow(W)$:
\begin{equation}\label{eq_det_combined}
  \begin{split}
  &\det \left( \tilde{M}_\uparrow(W)\, \tilde{M}^T_\downarrow(W) \right)
    = \\
    &=\sum_{p \in S_n} (-1)^{\epsilon(p)} \prod_{j=1}^n
    \sum_{l=1}^n \;(\tilde{M}_\uparrow(W))_{j l}\; (\tilde{M}_\downarrow(W))_{p_j l}  \\
  &= \sum_{p \in S_n} (-1)^{\epsilon(p)} \!\!\! \sum_{l_1, \ldots, l_n = 1}^n
    \prod_{j=1}^n \;(\tilde{M}_\uparrow(W))_{j l_j}\; (\tilde{M}_\downarrow(W))_{p_j l_j},
  \end{split}
\end{equation}
where $p=(p_1,\dots,p_n)$ is one out of a set $S_n$ of permutations of $(1,\dots,n)$ and $\epsilon(p)$ is its sign.
Let us remark that Eq.~\eqref{eq_det_combined} produces $n!\, n^n$ terms, whilst computing that same quantity from $\det (\tilde{M}_\uparrow(W)) \det
(\tilde{M}^T_\downarrow(W))$ only generates $(n!)^2$ terms.  The
reason for this discrepancy is a cancellation in the expression above whenever two
$l_j$'s have the same value: only those terms where all $l_j$'s are
different contribute.  As a consequence, expanding the determinant
of the matrix generates many unphysical diagrams.
The diagrammatic interpretation of the condition that
$l_j$'s' must all be different is that only diagrams where every
vertex carries exactly four propagators remain. If we impose that the diagonal
of $\tilde{M}_\uparrow(W)\, \tilde{M}^T_\downarrow(W)$ vanishes, part of the cancellation of unphysical diagrams does not occur.

\begin{figure}
\includegraphics[width=0.45\textwidth]{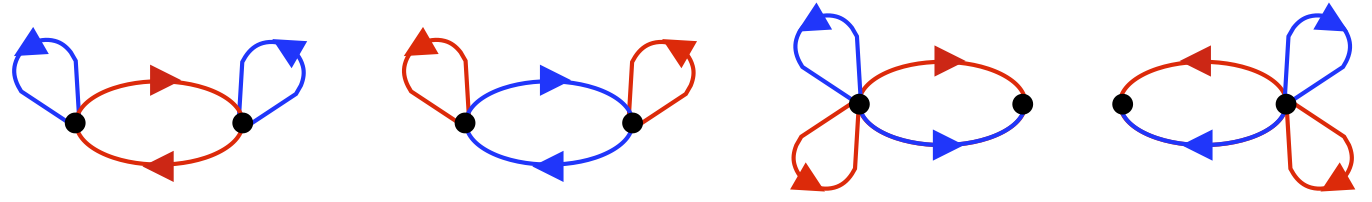}
\caption{Second-order diagrams for the choice of two local $U$ interaction vertices generated by the determinant of $\tilde{M}_\uparrow(W) \tilde{M}_\downarrow^T(W)$
when all diagonal elements are set to zero. The two rightmost diagrams are
unphysical.\label{fig_bad_diagrams}}
\end{figure}

In order to clarify the origin of unphysical diagrams, let us consider the case where we only have two local $U$ vertices. We eliminate from the $2\times 2$ matrix $\tilde{M}_\uparrow(W)\, \tilde{M}^{T}_\downarrow(W)$ the diagonal elements, and we compute the determinant of the resulting matrix:
\begin{equation}\label{eq_det_example}
  \det
  \begin{pmatrix}
    0 &
    \sum_{l_1} (\tilde{M}_\uparrow)_{1 l_1} (\tilde{M}_\downarrow)_{2 l_1} \\
    \sum_{l_2} (\tilde{M}_\uparrow)_{2 l_2} (\tilde{M}_\downarrow)_{1 l_2} &
    0
  \end{pmatrix},
\end{equation}
where we dropped the $W$ dependence of $\tilde{M}_\sigma(W)$.
In Fig.~\ref{fig_bad_diagrams} we draw the Feynman diagram interpretation of the terms resulting from the determinant expansion of Eq.~\eqref{eq_det_example}: we see that we have successfully eliminated ladder diagrams, while we have produced two unphysical diagrams and therefore obtained an incorrect expression.

\subsection{Elimination of unphysical diagrams}
\label{subsec_unphysical}

In order to eliminate unphysical diagrams, we introduce a matrix  $\tilde{M}(W,s)$, where $s:=\{s_1,\dots,s_n\}$,
which depends on artifical classical spin variables $s_j\in\{-1,1\}$, for $j\in\{1,\dots,n\}$:
\begin{equation}\label{eq_def_M_spin}
    \left(\tilde{M}(W,s)\right)_{jk}:=
    \sum_{l=1}^n s_l\;(\tilde{M}_\uparrow(W))_{jl}\;(\tilde{M}_\downarrow(W))_{kl}\;(1-\delta_{jk}).
\end{equation}
One has:
\begin{equation}\label{}
  \begin{split}
    &\frac{1}{2^n}\sum_{s_1,\dots,s_n\in\{-1,1\}} \!\!\! \det\tilde{M}(W,s) \prod_{j=1}^n s_j =\\
    & \sum_{p \in S_n} (-1)^{\epsilon(p)} \sum_{l\in S_n }
    \prod_{j=1}^n (\tilde{M}_\uparrow(W))_{j l_j} (\tilde{M}_\downarrow(W))_{p_j l_j}\,(1-\delta_{j,p_j}).
    \end{split}
\end{equation}

To summarize, in order to compute the contribution to the grand-canonical potential density $\Phi_G$ at order $n$ in $\xi$ of the action $S^{\text{F}}(\xi)$ from ~\eqref{eq_action_F_xi}, one needs to choose for each $\xi$ either the local vertex or the non-local vertex (see Eq.~\eqref{eq_action_I_xi} and Fig.~\ref{fig_vertices}). One builds the matrix $\tilde{M}(W,s)$ from Eq.~\eqref{eq_def_M_spin} and computes the sum of all connected and disconnected diagrams as:
\begin{equation}\label{}
a(W):=\frac{(-1)^{n+1}}{2^n} \sum_{s_1,\dots,s_n\in\{-1,1\}} \!\!\! \det\tilde{M}(W,s) \prod_{j=1}^n s_j.
\end{equation}
Then the recursive formula in Eq.~\eqref{eq_recursive_formula} can be used to eliminate disconnected diagrams and integrate each $c(W)$ over spacetime vertex positions as in Eq.~\eqref{eq_integral_G0P0}, and finally sum over the $2^n$ choices of the local/non-local vertices of Fig.~\ref{fig_vertices}.

\subsection{Integrating out of the non-local vertices}
\label{subsec_integration}

As described in the previous section, at order $n$ one has to sum over the $2^n$ choices of the vertices of Fig.~\ref{fig_vertices} as the two types of vertices have a different number of variables and cannot be sampled together. In the context of sampling individual Feynman-diagram topologies, in  Ref.~\onlinecite{deng} it was found advantageous to introduce an auxiliary non-local variable for the local $U$ vertex and sample both vertices at the same time. We choose a different strategy: We integrate out the $Y$ variable of the non-local vertex in Fig.~\ref{fig_vertices} in order to have the same number of variables for both vertices, thus making it possible to avoid the $2^n$ sum over all possible vertex combinations.

As a first step, we absorb the interaction vertices $U$ and $P_0^{\text{pp}}$ of Eq.~\eqref{eq_integral_G0P0} into the matrix $\tilde{M}(W,s)$ of Eq.~\eqref{eq_def_M_spin} and obtain the matrix:
\begin{equation}\label{eq_def_mathfrak_M}
  \begin{split}
    &\left(\mathcal{M}(W,s)\right)_{jk}:=\\
    &\sum_{l\in \{1,\dots,u\}\setminus\{j,k\}} s_l\;U\;G_0^\uparrow(X_j,X_l)\;G_0^\downarrow(X_k,X_l)+\\
    &+\sum_{l\in \{u+1,\dots,n\}} s_l\;P_0^{\text{pp}}(Y_l,X_l)\;G_0^\uparrow(X_j,Y_l)\;G_0^\downarrow(X_k,Y_l),
  \end{split}
\end{equation}
where $s_l\in\{-1,1\}$ as before.
We introduce:
\begin{equation}\label{}
  \mathcal{A}(W):=\frac{(-1)^{n+1}}{2^n} \sum_{s_1,\dots,s_n\in\{-1,1\}}  \!\!\! \det \;\mathcal{M}(W,s) \prod_{j=1}^n s_j.
\end{equation}
The selection of the $s_1\dots s_n$ component of the determinant guarantees that $U$ is chosen only once for each $l\in \{ 1,\dots,u\}$, and that $P_0^{\text{pp}}(Y_l,X_l)$ is chosen only once for each $l \in \{u+1,\dots,n\}$ (see \eqref{eq_integral_G0P0}). We then apply Eq.~\eqref{eq_recursive_formula} with the substitutions $a(W)\to \mathcal{A}(W)$ and $c(W)\to\mathcal{C}(W)$ in order to obtain the connected part $\mathcal{C}(W)$. We now rewrite Eq.~\eqref{eq_integral_G0P0} as
\begin{equation}\label{eq_integral_G0P0_mathfrak}
  \frac{\xi^n}{\beta L_x L_y n!}\int_{X_1,\dots,X_n,Y_{u+1},\dots,Y_n} \mathcal{C}(W).
\end{equation}
We stress that in order to obtain the complete $\xi^n$ contribution one has to sum over all $2^n$ choices of local and non-local vertices (see Fig.~\ref{fig_vertices}).

In order to consider directly the sum of all possible vertex choices, we introduce the function $\mathcal{L}_0^{\text{pp}}$, which consists out of a vertex to which two propagators are attached:
\begin{equation}\label{eq_lobster}
  \begin{split}
    &\mathcal{L}_0^{\text{pp}}(X',X'';X) := \\ & U\,G_0^\uparrow(X',X)\,G_0^\downarrow(X'',X)+\mathcal{L}_{0;\text{nl}}^{\text{pp}}(X',X'';X),
  \end{split}
\end{equation}
where
\begin{equation}\label{eq_L0nl}
  \begin{split}
&\mathcal{L}_{0;\text{nl}}^{\text{pp}}(X',X'';X):=\\&
\int_Y P_0^{\text{pp}}(Y,X)\,G_0^\uparrow(X',Y)\,G_0^\downarrow(X'',Y).
\end{split}
\end{equation}
With the introduction of the function $\mathcal{L}_{0;\text{nl}}^{\text{pp}}$ we can perform the integral over $Y_{u+1},\dots,Y_n$ of Eq.~\eqref{eq_integral_G0P0_mathfrak} exactly. Indeed, by re-introducing the set of vertices $V:=\{X_1,\dots,X_n\}$, we define the following matrix:
\begin{equation}\label{eq_def_mathfrak_M}
  \begin{split}
    &\left(\tilde{\mathcal{M}}(V,s)\right)_{jk}:=\\
    &\sum_{l\in \{1,\dots,u\}\setminus\{j,k\}} s_l\;U\;G_0^\uparrow(X_j,X_l)\;G_0^\downarrow(X_k,X_l)+\\
    &+\sum_{l\in \{u+1,\dots,n\}} s_l\;\mathcal{L}_{0;\text{nl}}^{\text{pp}}(X_j,X_k;X_l),
  \end{split}
\end{equation}
and the corresponding:
\begin{equation}\label{eq_def_mathfrak_a}
  \tilde{\mathcal{A}}(V):=\frac{(-1)^{n+1}}{2^n}
\sum_{s_1,\dots,s_n\in\{-1,1\}} \!\!\! \det \;\tilde{\mathcal{M}}(V,s) \prod_{j=1}^n s_j.
\end{equation}
One can see that Eq.~\eqref{eq_integral_G0P0_mathfrak}, after the application of Eq.~\eqref{eq_recursive_formula} with the substitutions $a(V)\to\tilde{\mathcal{A}}(V)$ and $c(V)\to\tilde{\mathcal{C}}(V)$, becomes
\begin{equation}\label{eq_integral_G0P0_t}
  \frac{\xi^n}{\beta L_x L_y n!}\int_{X_1,\dots,X_n} \tilde{\mathcal{C}}(V).
\end{equation}
One, however, still has to sum over all possible choices of local/non-local vertices of Fig.~\ref{fig_vertices}.

The final formulation consists of considering directly the sum over all possible choices of interaction vertices.
To achieve this, we introduce the following matrix:
\begin{equation}\label{eq_def_mathfrak_M}
    \left(\bar{\mathcal{M}}(V,s)\right)_{jk}:=\sum_{l=1}^n s_l\;\bar{\mathcal{L}}_0^{\text{pp}}(X_j,X_k;X_l),
\end{equation}
where we define
\begin{equation}\label{}
  \bar{\mathcal{L}}_0^{\text{pp}}(X_j,X_k;X_l):=\left\{\begin{array}{ll}\mathcal{L}_0^{\text{pp}}(X_j,X_k;X_l)\quad & \text{for}\; j\neq l\wedge k\neq l\\
 \mathcal{L}_{0;\text{nl}}^{\text{pp}}(X_j,X_k;X_l) & \text{otherwise}\end{array}\right.
\end{equation}
We can now define $\bar{\mathcal{A}}(V)$ from Eq.~\eqref{eq_def_mathfrak_a} with the substituion $\tilde{\mathcal{A}}(W)\to \bar{\mathcal{A}}(V)$ and $\tilde{\mathcal{M}}(W,s)\to \bar{\mathcal{M}}(V,s)$. We also define $\bar{\mathcal{C}}(V)$ from Eq.~\eqref{eq_recursive_formula} with the substitutions $c(V)\to \bar{\mathcal{C}}(V)$ and $a(V)\to \bar{\mathcal{A}}(V)$. We can finally write the expression for the order $n$ contribution to the grand-canonical potential density $\Phi_G$ as:
\begin{equation}\label{}
  \phi_{0;n}^{\text{pp}}=  \frac{1}{\beta L_x L_y n!}\int_{X_1,\dots,X_n} \bar{\mathcal{C}}(V).
\end{equation}

\subsection{Computational cost and numerical implementation}
\label{subsec_complexity}

In this section we briefly discuss the computational cost and the spectral compression of the function $\mathcal{L}_{0;\text{nl}}^{\text{pp}}$.

The computational cost of computing determinants, and summing over spin variables (see Eq.~\eqref{eq_def_mathfrak_a}), at order $n$ in $\xi$, is proportional to
\begin{equation}\label{}
 \sum_{k=0}^n\,2^k\,\left(\begin{array}{c} n \\ k\end{array}\right) k^3\sim \mathcal{O}(n^3\, 3^n),
\end{equation}
where $2^k$ comes from the spin trace, $\left(\begin{array}{c}n \\ k\end{array}\right)$ is the number of subsets of $V$ with cardinality $k$, and $k^3$ is roughly the cost of computing a $k\times k$ determinant.
  We note that this cost cannot be alleviated in this situation by the fast principal minor
algorithm~\cite{griffin2006minors}, generally used in bare interaction CDet, due to the fact that
minors no longer correspond to determinants for subsets of the full set.
The $3^n$ computational cost of applying the recursive formula \cite{cdet} (or, alternatively $n^2 2^n$ \cite{koivisto}), Eq.~\eqref{eq_recursive_formula} , is negligeable compared to the aforementioned cost. The exponential scaling of the algorithm means that the resulting computational scaling of the inverse error with computational time is polynomial inside the radius of convergence~\cite{rr_epl}.

We consider now the numerical compression and storage of the function
\begin{equation}
\mathcal{L}_{0;\text{nl}}^{{\text{pp}}}(X',X'';X)=\mathcal{L}_{0;\text{nl}}^{{\text{pp}}}(X'-X,X''-X),
\end{equation}
where we used translation invariance. Without loss of generality, we can therefore suppose $X=(\mathbf{r},\tau)=(\mathbf{0},0)$. As \begin{equation}
\mathcal{L}_{0;\text{nl}}^{\text{pp}}(X',X'') = \mathcal{L}_{0;\text{nl}}^{\text{pp}}(X'',X'),
\end{equation}
we can suppose that if $X'=(\mathbf{r}',\tau')$ and $X''=(\mathbf{r}'',\tau'')$, then $0\le \tau'\le \tau''\le \beta$. We can then write:
\begin{equation}\label{eq_chebyshev_L}
  \begin{split}
    &\mathcal{L}_{0;\text{nl}}^{{\text{pp}}}((\mathbf{r}',\tau'),(\mathbf{r}'',\tau''))=\\
    &=\sum_{\mathbf{r}}\int_0^{\tau'} d\tau\;
    P_0^{\text{pp}}(\mathbf{r},\tau)\,\times\\
    &
    \times G_0^\uparrow(\mathbf{r}'-\mathbf{r},\tau'-\tau)\,
    G_0^\downarrow(\mathbf{r}''-\mathbf{r},\tau''-\tau)\\
    &-\sum_{\mathbf{r}}\int_{\tau'}^{\tau''} d\tau\;
    P_0^{\text{pp}}(\mathbf{r},\tau)\,\times\\
    &\times G_0^\uparrow(\mathbf{r}'-\mathbf{r},\beta+\tau'-\tau)\,
    G_0^\downarrow(\mathbf{r}''-\mathbf{r},\tau''-\tau)\\
    &+\sum_{\mathbf{r}}\int_{\tau''}^{\beta} d\tau\;
    P_0^{\text{pp}}(\mathbf{r},\tau)\,\times\\
    &
    \times G_0^\uparrow(\mathbf{r}'-\mathbf{r},\beta+\tau'-\tau)\,
    G_0^\downarrow(\mathbf{r}''-\mathbf{r},\beta+\tau''-\tau).
    \end{split}
\end{equation}
We expand $\mathcal{L}_{0;\text{nl}}^{\text{pp}}$ as defined in Eq.~\eqref{eq_chebyshev_L} in two-dimensional Chebyshev polynomials for imaginary times $\tau',\tau''\in [0,\beta]$, and for each value of the lattice sites $\mathbf{r}$ and $\mathbf{r}'$. For the purpose of Chebyshev interpolation, it is important to use Eq.~\eqref{eq_chebyshev_L} for $\tau''<\tau'$ as well, with the imaginary-time analytic continuation of $G_0^\sigma$, as this guarantees a smooth function of $\tau'$ and $\tau''$, which implies a very fast convergence of our spectral representation. The physical result can be obtained by using the symmetry between $X'$ and $X''$ and only evaluating the expression when $\tau'<\tau''$.

In practice, we use a $5\times 5$ grid for both $\mathbf{r}'$ and $\mathbf{r}''$, and we store the Chebyshev polynomial representation of $\mathcal{L}_{0;\text{nl}}^{\text{pp}}$ inside this grid. As we deal with connected diagrams, the Monte Carlo sampling rarely goes outside this grid, and in those cases when it does one can afford to compute the $\mathcal{L}_{0;\text{nl}}^{\text{pp}}$ function on the fly.

\section{$\mathbf{G_1\,P_1^{\text{pp}}}$ expansion}
\label{sec_G1P1}

\begin{figure}
  \includegraphics[width=0.49\textwidth]{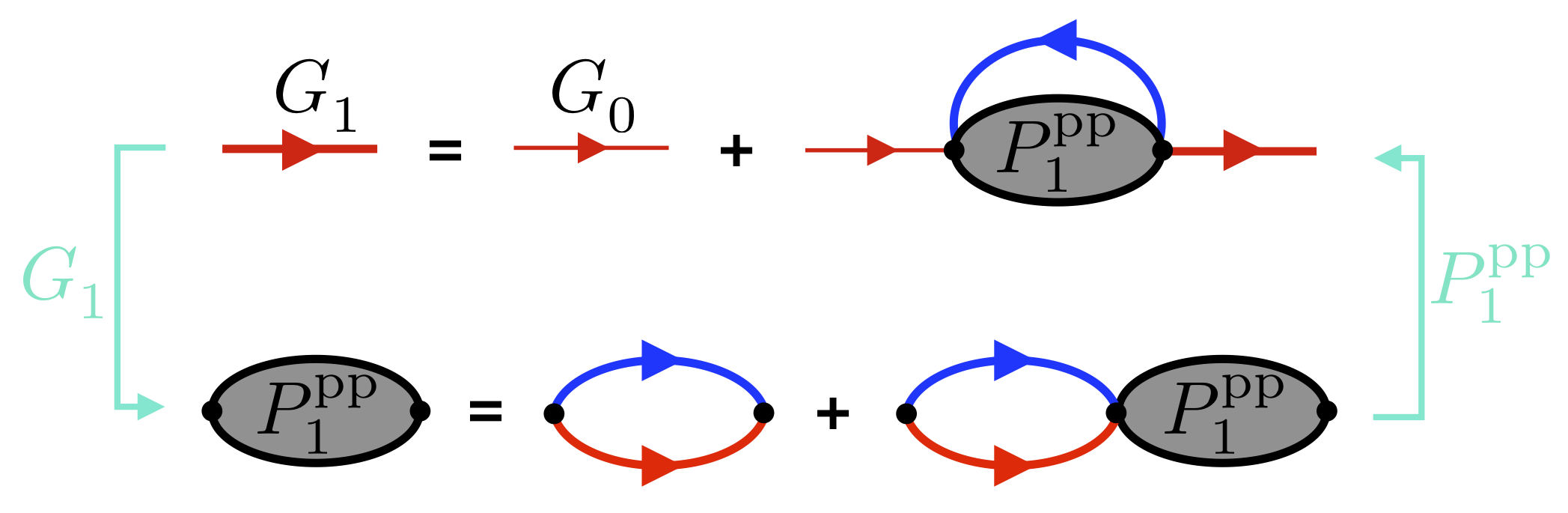}
\caption{Self-consistent loop for obtaining the first-order semi-bold propagator $G_1$ and first-order semi-bold vertex $P_1^{\text{pp}}$. The first equation can identically be written with spin-colors inverted.\label{fig_selfconsistent_loop}}
\end{figure}

One can take the diagrammatic renormalization one step further by
self-consistently determining the non-local Hartree term, which results in a
diagrammatic expansion denoted as the ``first-order semibold'' expansion
in Ref.~\onlinecite{shifted_action},
and which we call $G_1 P_1^{\text{pp}}$ expansion in what follows.
We define the following set of equations for $G_1$ and $P_1^{\text{pp}}$:
\begin{equation}\label{eq_selfconsistent_loop}
  \begin{split}
   &G_1^{\sigma}(X,X') = G_0^{\sigma}(X,X')+\\
    &+ \int_{Y,Y'}G_0^{\sigma}(X,Y)\, P_1^{\text{pp}}(Y,Y')\, G_1^{\bar{\sigma}}(Y',Y)\, G_1^{\sigma}(Y',X') \\
 &P_1^{\text{pp}}(K) = U^2\, \tilde{P}_1^{\text{pp}}(K) + U\, \tilde{P}_1^{\text{pp}}(K)\, P_1^{\text{pp}}(K),
 \end{split}
\end{equation}
where
\begin{equation}
  \tilde{P}^{\text{pp}}_1(X,Y) := -G_1^\uparrow(X,Y) G_1^\downarrow(X,Y).
\end{equation}
We also provide the diagrammatic interpretation of this set of equations in Fig.~\ref{fig_selfconsistent_loop}.

\subsection{Shifted action}
\label{subsec_G1P1_action}

In this section, we present the shifted-action expressions for the $G_1 P_1^{\text{pp}}$ expansion, which is equivalent to an expansion in powers of $\xi$ with this formalism:
\begin{equation}\label{eq_action_F_0_xi_1}
  \begin{split}
S^{\text{F}}_{1}(\xi)&:=  -\sum_{\sigma}
\int_X \bar{\psi}_\sigma(X) \left((G_{1}^\sigma)^{-1}\psi_\sigma\right)(X)\\
&-\xi\, U\;\sum_{\sigma}n_1^{\bar{\sigma}}\int_X \left(\bar{\psi}^{\sigma}\psi^{\sigma}\right)(X)\\
& -\xi \sum_{\sigma}\int_{X,Y} \bar{\psi}_\sigma(Y)\, P_{1}^{\text{pp}}(Y,X)\,G_1^{\bar{\sigma}}(X,Y)\,\psi_\sigma(X)\\
  \end{split}
\end{equation}
where $n_1^{\sigma}:=G_1^\sigma(\mathbf{r}=\mathbf{0},\tau=0^{-})$,
\begin{equation}\label{eq_action_HS_0_xi_1}
  \begin{split}
    S^{\text{HS}}_{1}(\xi)&:= \frac{1}{U}\int_X (\bar{\eta}\eta)(X)+\\
    &+(1-\xi)\int_{Y,X} \bar{\eta}(Y)\,(G_1^\uparrow G_1^\downarrow)(Y,X)\,\eta(X)
    \end{split}
\end{equation}
\begin{align}\label{eq_action_R_1}
  S^{\text{HS}}_{I}(\xi) := i\sqrt{\xi}\int_X \bar{\eta}(X)\;(\psi_\downarrow \psi_\uparrow)(X)+h.c.
\end{align}
and the shifted action is
\begin{equation}
  S^{\text{HS}}(\xi):=S^{\text{F}}_{1}(\xi)+S^{\text{HS}}_{1}(\xi)+S^{\text{HS}}_{I}(\xi).
\end{equation}

It is then possible to integrate out the Hubbard-Stratonovich field $\eta$ to obtain the analogous of Eq.~\eqref{eq_action_F_xi}.

\subsection{Feynman-diagrammatic definition}
\label{subsec_G1P1_feynman}

\begin{figure}
  \includegraphics[width=0.49\textwidth]{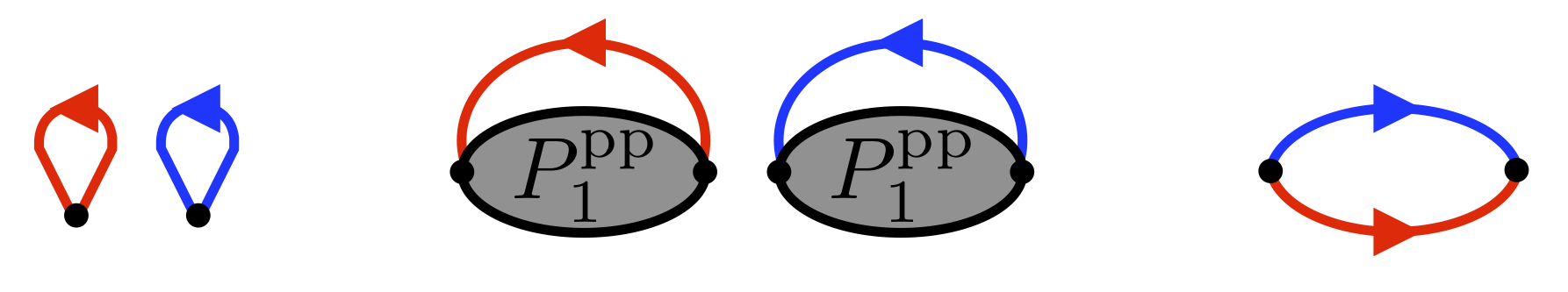}
  \caption{Feynman-diagram insertions that cannot appear in the particle-particle $G_1 P^{\text{pp}}_1$ expansion. This includes the local tadpoles (on the left), the non-local tadpoles (in the center) and particle-particle bubble (on the right).
    \label{fig_G1P1_removed}}
\end{figure}

\begin{figure}
  \includegraphics[width=0.49\textwidth]{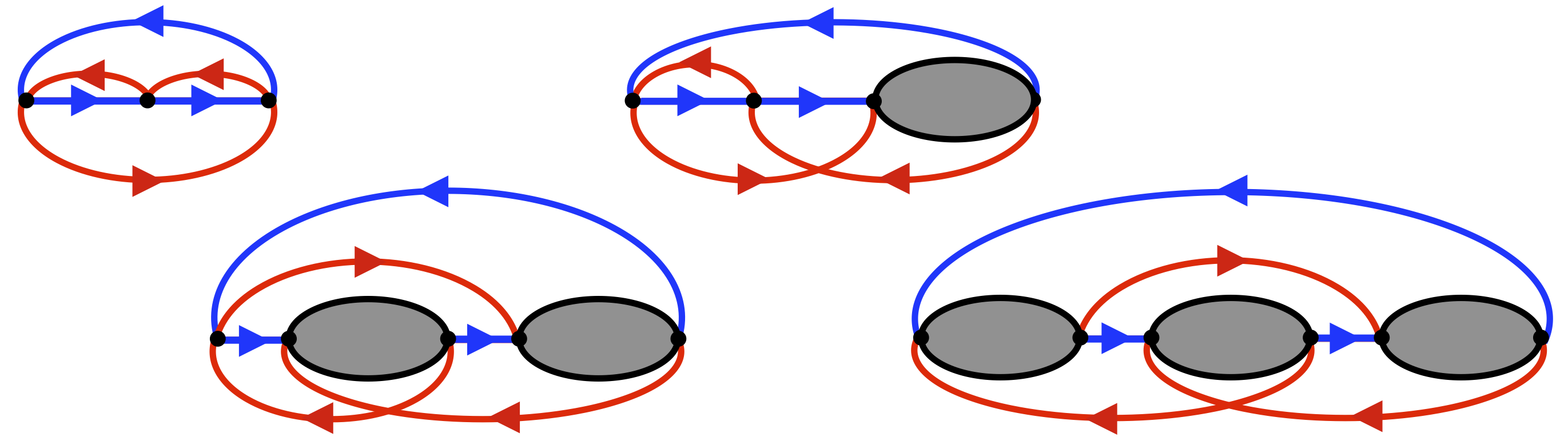}
\caption{All third order Feynman diagrams of the $G_1 P_1^{\text{pp}}$ expansion. No first and second order diagrams exist in this expansion.\label{fig_GP}}
\end{figure}

In Fig.~\ref{fig_G1P1_removed} we draw the diagram insertions which are forbidden in this expansion. In Fig.~\ref{fig_GP}, we present the $G_1\,P_1^{\text{pp}}$ diagrammatic expansion for the grand-canonical potential density $\Phi_G$ up to third order. Note that the first two orders in this expansion contain no diagrams.

\subsection{Connected Determinant Diagrammatic Monte Carlo}
\label{subsec_G1P1_cdet}

In order to consider the $G_1 P_1^{\text{pp}}$ expansion within the CDet framework, one needs to take into account the following modifications to the discussion for the $G_0P_0^{\text{pp}}$ expansion: One needs to additionally eliminate all self-loops from the matrix $\tilde{M}(W)$, previously defined in equation~\eqref{eq_def_tilde_M}:
\begin{equation}\label{eq_def_tilde_M_new}
  (\tilde{M}_\sigma(W))_{jk}:=(1-\delta_{jk})\;
  (M_\sigma(W))_{jk},
\end{equation}
and to substitute $G_0^\sigma$ with $G_1^\sigma$ and $P_0^{\text{pp}}$ with $P_1^{\text{pp}}$. Note that $G_1^\sigma$ and $P_1^{\text{pp}}$ are computed by the self-consistent evalution of Eq.~\eqref{eq_selfconsistent_loop}, as displayed in Fig.~\ref{fig_selfconsistent_loop}, before the start of the Monte Carlo loop.
\begin{figure}
  \includegraphics[width=0.45\textwidth]{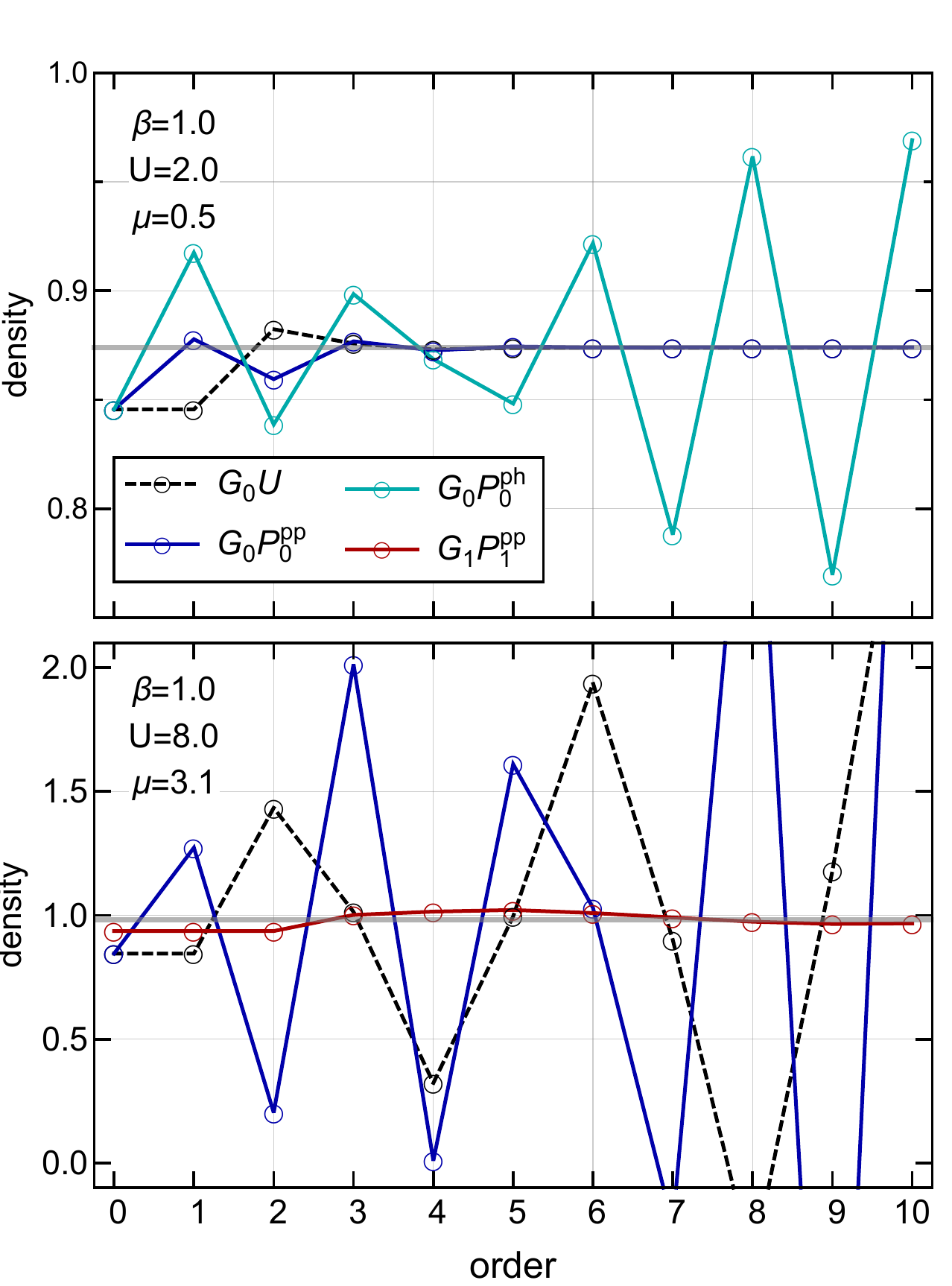}
\caption{Partial sum of the density series for two interaction values $U=2$ (top) and $U=8$ (bottom) of the Hubbard atom computed from different diagrammatic expansions. The exact solutions are given by gray lines. \label{fig_hubbard_atom}}
\end{figure}

\section{Numerical results}
\label{sec_results}

We proceed by showcasing numerical results obtained for the density using the technique we have introduced for the $G_0 P_0^{\text{pp}}$, $G_0 P_0^{\text{ph}}$, and $G_1 P_1^{\text{pp}}$ expansions. First, we would like to stress that, for all of the renormalized expansions considered in this work,
the series in $\xi$ (see Eq.~\eqref{grand_potential_series}) only correspond to the original physical model
when evaluated at $\xi = 1$. This is in contrast with the usual bare
interaction series which gives a physical result for some value of the chemical potential and interaction strength
for all $\xi$. This property of the renormalized series turns out to be an
advantage: one can avoid the appearance of singularities on the negative
real axis as the series does not need to be physical for negative interaction strengths.
As a result, the series can have a radius of
convergence which includes the physical value of interest. In comparison, the series resulting from a second-order one-particle renormalization, as introduced in Ref.~\onlinecite{rossi2020renormalized}, yield a physical result at both $\xi = 1$ and $\xi = -1$, and is thus affected by the negative real axis singularities.

\subsection{Hubbard Atom}
\label{subsec_atom}

In Fig.~\ref{fig_hubbard_atom}, we present benchmark results for the density of the Hubbard atom at weak interactions ($U=2$, upper panel) computed for the bare interaction,  $G_0 U$ (bare interaction) expansion as well as the $G_0 P_0^{\text{pp}}$ and $G_0 P_0^{\text{ph}}$ expansions, and we compare to the exact analytical result. We see that both the $G_0 U$ and the $G_0 P_0^{\text{pp}}$ series converge to the exact result within a few orders whilst the $G_0 P_0^{\text{ph}}$ series is divergent.
At strong interactions ($U=8$, lower panel) we see that both the $G_0 U$ and the $G_0 P_0^{\text{pp}}$ series are strongly oscillating and diverging. However, the $G_1 P_1^{\text{pp}}$ series turns out to be converging quickly and is easily resummed to the exact result.

\begin{figure}
  \includegraphics[width=0.45\textwidth]{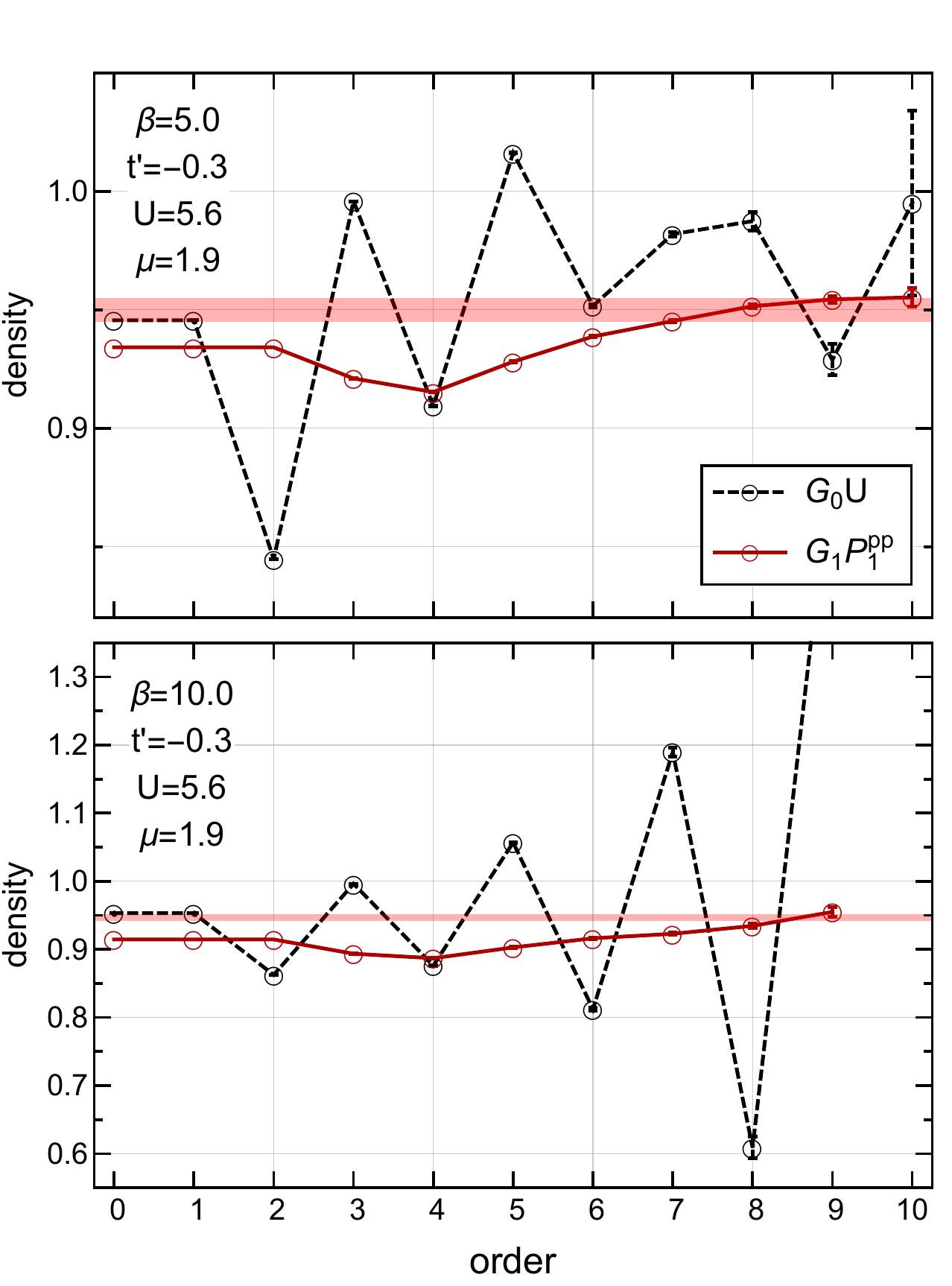}
\caption{Partial sum of the density series at two temperatures $\beta=5$ (top) and $\beta=10$ (bottom) computed from different expansions. The horizontal bands show extrapolated results. \label{fig_Wei_point}}
\end{figure}

\subsection{Two-dimensional Hubbard model}
\label{subsec_2d_hubbard}

We now present numerical results obtained in the two-dimensional Hubbard model,
away from half-filling and with particle-hole asymmetry ($t'=-0.3$, $U=5.6$, $\beta=\{ 5,
10 \}$). The limit of the computation of Ref.~\onlinecite{wu_controlling} was $\beta=5$.
At both evaluated temperatures, the
$G_1 P_1^{\text{pp}}$ series for the density shows a remarkably better convergence than the
$G_0 U$ bare interaction series.
At $\beta = 5$, the
$G_1 P_1^{\text{pp}}$ series is clearly convergent and easily resummable.
At $\beta =10$, the series also seems convergent and can be resummed, however, an
additional oscillatory behavior appears at higher orders,
hinting at the appearance of poles in the complex plane near the negative real axis.

Another advantage of using renormalized vertices is the reduced real-space
spread of Feynman diagrams. In the bare-interaction CDet algorithm, as the perturbation
order grows, the sampled diagrams extend wider in real space. As a consequence, the effective configuration space to sample is larger and the variance increases, making
it difficult to compute large perturbation orders. In a generic situation, the
diagrams with the greatest spread are of the form of a chain of tadpoles. However, if the perturbation theory is constructed around mean-field, such as in our case, tadpole insertions vanish and the leftover diagrams are more concentrated yielding a smaller variance. An inspection of the most spread diagrams in that case shows that they are
made of chains of bubble diagrams. These diagrams, too, vanish for the above described expansions, thus further decreasing the extent and variance and allowing for the computation of higher perturbation orders. Naturally, as temperature is
lowered further, other classes of diagrams eventually start to spread
and it becomes difficult to compute large perturbation orders with great
accuracy.

\section{Conclusion}
\label{sec_conclusion}

We have presented an efficient and systematic way of computing perturbative expansions
based on one-loop renormalized interaction vertices using determinants.
We have considered the diagrammatic expansion around the random-phase approximation
in both the particle-particle and the
particle-hole channel, and have shown how the two-body long-range interaction can
be integrated out to yield an effective zero-range interaction with several
computational advantages.
This was achieved by using a determinantal formalism and the spacetime
representation, within the framework of Connected Determinant Monte Carlo~\cite{cdet}.
The computational cost, while bigger than the corresponding bare-interaction algorithm,
is still exponential in diagram order, resulting in a overall polynomial scaling of
the errorbar as a function of computational time inside the radius of convergence~\cite{rr_epl}.
We have further presented benchmark calculations in the two-dimensional Hubbard model away from half-filling, showing that with the technique we introduced in this work is able
to compute about 10 expansion order coefficients, and that the resulting series is
much better behaved than the original bare-interaction expansion series. From
a computational point of view, we have also witnessed an improvement to the Monte Carlo variance.

Summing up, we have shown that
expansions based on renormalized interaction
vertices are an interesting and practical
direction for unbiased diagrammatic calculations,
and how it is possible
to efficiently and systematically
implement them using the determinantal formalism,
thus opening new opportunities
for quantum many-body simulations.
As a future perspective, this method can be applied to the electron gas, where
it can prove useful in order to avoid the divergencies of RPA
bubble diagrams and work directly in the thermodynamic limit.
It would also be interesting to study whether vertex renormalized series can
be used to understand the onset of superconductivity and/or stripes in the Hubbard
model at low temperatures, where the bare interaction series is difficult to
resum.

We thank F.~Werner, K.~Van~Houcke and T.~Ohgoe for valuable discussions. This work was granted access to the HPC resources of TGCC and IDRIS under the allocations A0070510609 and A0050510609 attributed by GENCI (Grand Equipement National de Calcul Intensif). It has also been supported by the Simons Foundation within the Many Electron Collaboration
framework. The Flatiron Institute is a division of the Simons Foundation.

\bibliographystyle{ieeetr}
\bibliography{main_biblio}

\begin{thebibliography}{10}

\bibitem{leblanc2015solutions}
J.~LeBlanc, A.~E. Antipov, F.~Becca, I.~W. Bulik, G.~K.-L. Chan, C.-M. Chung,
  Y.~Deng, M.~Ferrero, T.~M. Henderson, C.~A. Jim{\'e}nez-Hoyos, {\em et~al.},
  ``Solutions of the two-dimensional hubbard model: benchmarks and results from
  a wide range of numerical algorithms,'' {\em Physical Review X}, vol.~5,
  no.~4, p.~041041, 2015.

\bibitem{schafer2020tracking}
T.~Schäfer, N.~Wentzell, F.~Šimkovic IV, Y.-Y. He, C.~Hille, M.~Klett, C.~J.
  Eckhardt, B.~Arzhang, V.~Harkov, F.-M.~L. Régent, A.~Kirsch, Y.~Wang, A.~J.
  Kim, E.~Kozik, E.~A. Stepanov, A.~Kauch, S.~Andergassen, P.~Hansmann,
  D.~Rohe, Y.~M. Vilk, J.~P.~F. LeBlanc, S.~Zhang, A.~M.~S. Tremblay,
  M.~Ferrero, O.~Parcollet, and A.~Georges, ``Tracking the footprints of spin
  fluctuations: A multi-method, multi-messenger study of the two-dimensional
  hubbard model,'' {\em arXiv preprint arXiv:2006.10769}, 2020.

\bibitem{jaksch1998cold}
D.~Jaksch, C.~Bruder, J.~I. Cirac, C.~W. Gardiner, and P.~Zoller, ``Cold
  bosonic atoms in optical lattices,'' {\em Physical Review Letters}, vol.~81,
  no.~15, p.~3108, 1998.

\bibitem{Bloch_review_2005}
I.~Bloch, ``Ultracold quantum gases in optical lattices,'' {\em Nature
  Physics}, vol.~1, pp.~23 EP --, 10 2005.

\bibitem{kohl2005fermionic}
M.~K{\"o}hl, H.~Moritz, T.~St{\"o}ferle, K.~G{\"u}nter, and T.~Esslinger,
  ``Fermionic atoms in a three dimensional optical lattice: Observing fermi
  surfaces, dynamics, and interactions,'' {\em Physical Review Letters},
  vol.~94, no.~8, p.~080403, 2005.

\bibitem{lewenstein2007ultracold}
M.~Lewenstein, A.~Sanpera, V.~Ahufinger, B.~Damski, A.~Sen, and U.~Sen,
  ``Ultracold atomic gases in optical lattices: mimicking condensed matter
  physics and beyond,'' {\em Advances in Physics}, vol.~56, no.~2,
  pp.~243--379, 2007.

\bibitem{jordens2008mott}
R.~J{\"o}rdens, N.~Strohmaier, K.~G{\"u}nter, H.~Moritz, and T.~Esslinger, ``A
  mott insulator of fermionic atoms in an optical lattice,'' {\em Nature},
  vol.~455, no.~7210, pp.~204--207, 2008.

\bibitem{schneider2008metallic}
U.~Schneider, L.~Hackerm{\"u}ller, S.~Will, T.~Best, I.~Bloch, T.~Costi,
  R.~Helmes, D.~Rasch, and A.~Rosch, ``Metallic and insulating phases of
  repulsively interacting fermions in a 3d optical lattice,'' {\em Science},
  vol.~322, no.~5907, pp.~1520--1525, 2008.

\bibitem{hulet2015antiferromagnetism}
R.~G. Hulet, P.~M. Duarte, R.~A. Hart, and T.-L. Yang, ``Antiferromagnetism
  with ultracold atoms,'' in {\em Laser Spectroscopy}, pp.~43--49, 2016.

\bibitem{greif2015formation}
D.~Greif, G.~Jotzu, M.~Messer, R.~Desbuquois, and T.~Esslinger, ``Formation and
  dynamics of antiferromagnetic correlations in tunable optical lattices,''
  {\em Physical Review Letters}, vol.~115, no.~26, p.~260401, 2015.

\bibitem{parsons2016site}
M.~F. Parsons, A.~Mazurenko, C.~S. Chiu, G.~Ji, D.~Greif, and M.~Greiner,
  ``Site-resolved measurement of the spin-correlation function in the
  fermi-hubbard model,'' {\em Science}, vol.~353, no.~6305, pp.~1253--1256,
  2016.

\bibitem{cheuk2016observation}
L.~W. Cheuk, M.~A. Nichols, K.~R. Lawrence, M.~Okan, H.~Zhang, E.~Khatami,
  N.~Trivedi, T.~Paiva, M.~Rigol, and M.~W. Zwierlein, ``Observation of spatial
  charge and spin correlations in the 2d fermi-hubbard model,'' {\em Science},
  vol.~353, no.~6305, pp.~1260--1264, 2016.

\bibitem{greiner2017}
A.~Mazurenko, C.~S. Chiu, G.~Ji, M.~F. Parsons, M.~Kan{\'a}sz-Nagy, R.~Schmidt,
  F.~Grusdt, E.~Demler, D.~Greif, and M.~Greiner, ``A cold-atom fermi--hubbard
  antiferromagnet,'' {\em Nature}, vol.~545, pp.~462 EP --, 05 2017.

\bibitem{nichols2019spin}
M.~A. Nichols, L.~W. Cheuk, M.~Okan, T.~R. Hartke, E.~Mendez, T.~Senthil,
  E.~Khatami, H.~Zhang, and M.~W. Zwierlein, ``Spin transport in a mott
  insulator of ultracold fermions,'' {\em Science}, vol.~363, no.~6425,
  pp.~383--387, 2019.

\bibitem{hartke2020measuring}
T.~Hartke, B.~Oreg, N.~Jia, and M.~Zwierlein, ``Measuring total density
  correlations in a fermi-hubbard gas via bilayer microscopy,'' {\em arXiv
  preprint arXiv:2003.11669}, 2020.

\bibitem{ProkSvistFrohlichPolaron}
N.~V. Prokof'ev and B.~V. Svistunov, ``Polaron problem by diagrammatic quantum
  monte carlo,'' {\em Phys. Rev. Lett.}, vol.~81, p.~2514, 1998.

\bibitem{ProkofevSvistunovPolaronLong}
N.~Prokof'ev and B.~Svistunov, ``Bold diagrammatic monte carlo: A generic
  sign-problem tolerant technique for polaron models and possibly interacting
  many-body problems,'' {\em Phys. Rev. B}, vol.~77, p.~125101, 2008.

\bibitem{van2010diagrammatic}
K.~Van~Houcke, E.~Kozik, N.~Prokof’ev, and B.~Svistunov, ``Diagrammatic monte
  carlo,'' {\em Physics Procedia}, vol.~6, pp.~95--105, 2010.

\bibitem{kris_felix}
K.~Van~Houcke, F.~Werner, E.~Kozik, N.~Prokof'ev, B.~Svistunov, M.~Ku,
  A.~Sommer, L.~Cheuk, A.~Schirotzek, and M.~Zwierlein, ``Feynman diagrams
  versus fermi-gas feynman emulator,'' {\em Nature Physics}, vol.~8, no.~5,
  pp.~366--370, 2012.

\bibitem{deng}
Y.~Deng, E.~Kozik, N.~V. Prokof'ev, and B.~V. Svistunov, ``Emergent bcs regime
  of the two-dimensional fermionic hubbard model: Ground-state phase diagram,''
  {\em EPL}, vol.~110, no.~5, 2015.

\bibitem{vsimkovic2019superfluid}
F.~Šimkovic IV, Y.~Deng, and E.~Kozik, ``Superfluid ground-state phase diagram
  of the $2 d $ hubbard model in the emergent bcs regime,'' {\em arXiv},
  pp.~arXiv--1912, 2019.

\bibitem{wu_controlling}
W.~Wu, M.~Ferrero, A.~Georges, and E.~Kozik, ``Controlling feynman diagrammatic
  expansions: Physical nature of the pseudogap in the two-dimensional hubbard
  model,'' {\em Phys. Rev. B}, vol.~96, p.~041105, Jul 2017.

\bibitem{kun_chen}
K.~Chen and K.~Haule, ``A combined variational and diagrammatic quantum monte
  carlo approach to the many-electron problem,'' {\em Nature Communications},
  vol.~10, no.~2725, 2019.

\bibitem{gull_inchworm}
I.~Krivenko, J.~Kleinhenz, G.~Cohen, and E.~Gull, ``Dynamics of kondo voltage
  splitting after a quantum quench,'' {\em Phys. Rev. B}, vol.~100, p.~201104,
  Nov 2019.

\bibitem{vucicevic2019real}
J.~Vucicevic and M.~Ferrero, ``Real-frequency diagrammatic monte carlo at
  finite temperature,'' {\em arXiv preprint arXiv:1908.11826}, 2019.

\bibitem{taheridehkordi2019algorithmic}
A.~Taheridehkordi, S.~Curnoe, and J.~LeBlanc, ``Algorithmic matsubara
  integration for hubbard-like models,'' {\em Physical Review B}, vol.~99,
  no.~3, p.~035120, 2019.

\bibitem{taheridehkordi2019optimal}
A.~Taheridehkordi, S.~Curnoe, and J.~LeBlanc, ``Optimal grouping of arbitrary
  diagrammatic expansions via analytic pole structure,'' {\em arXiv preprint
  arXiv:1911.11129}, 2019.

\bibitem{cdet}
R.~Rossi, ``Determinant diagrammatic monte carlo algorithm in the thermodynamic
  limit,'' {\em Phys. Rev. Lett.}, vol.~119, p.~045701, Jul 2017.

\bibitem{fedor_sigma}
F.~Šimkovic IV and E.~Kozik, ``Determinant monte carlo for irreducible feynman
  diagrams in the strongly correlated regime,'' {\em Phys. Rev. B}, vol.~100,
  p.~121102, Sep 2019.

\bibitem{alice_michel}
A.~Moutenet, W.~Wu, and M.~Ferrero, ``Determinant monte carlo algorithms for
  dynamical quantities in fermionic systems,'' {\em Phys. Rev. B}, vol.~97,
  p.~085117, Feb 2018.

\bibitem{rr_sigma}
R.~Rossi, ``Direct sampling of the self-energy with connected determinant monte
  carlo,'' {\em arXiv:1802.04743}, 2018.

\bibitem{rr_epl}
R.~Rossi, N.~Prokof'ev, B.~Svistunov, K.~Van~Houcke, and F.~Werner,
  ``Polynomial complexity despite the fermionic sign,'' {\em EPL}, vol.~118,
  no.~1, 2017.

\bibitem{olivier}
R.~E.~V. Profumo, C.~Groth, L.~Messio, O.~Parcollet, and X.~Waintal, ``Quantum
  monte carlo for correlated out-of-equilibrium nanoelectronic devices,'' {\em
  Phys. Rev. B}, vol.~91, p.~245154, Jun 2015.

\bibitem{corentin}
C.~Bertrand, S.~Florens, O.~Parcollet, and X.~Waintal, ``Reconstructing
  nonequilibrium regimes of quantum many-body systems from the analytical
  structure of perturbative expansions,'' {\em Phys. Rev. X}, vol.~9,
  p.~041008, Oct 2019.

\bibitem{kid_gull_cohen}
A.~Boag, E.~Gull, and G.~Cohen, ``Inclusion-exclusion principle for many-body
  diagrammatics,'' {\em Phys. Rev. B}, vol.~98, p.~115152, Sep 2018.

\bibitem{moutenet2019cancellation}
A.~Moutenet, P.~Seth, M.~Ferrero, and O.~Parcollet, ``Cancellation of vacuum
  diagrams and the long-time limit in out-of-equilibrium diagrammatic quantum
  monte carlo,'' {\em Physical Review B}, vol.~100, no.~8, p.~085125, 2019.

\bibitem{mavcek2020quantum}
M.~Ma\ifmmode~\check{c}\else \v{c}\fi{}ek, P.~T. Dumitrescu, C.~Bertrand,
  B.~Triggs, O.~Parcollet, and X.~Waintal, ``Quantum quasi-monte carlo
  technique for many-body perturbative expansions,'' {\em Phys. Rev. Lett.},
  vol.~125, p.~047702, Jul 2020.

\bibitem{kozik2010diagrammatic}
E.~Kozik, K.~Van~Houcke, E.~Gull, L.~Pollet, N.~Prokof'ev, B.~Svistunov, and
  M.~Troyer, ``Diagrammatic monte carlo for correlated fermions,'' {\em EPL
  (Europhysics Letters)}, vol.~90, no.~1, p.~10004, 2010.

\bibitem{fedor_hf}
F.~Šimkovic IV, J.~P.~F. LeBlanc, A.~J. Kim, Y.~Deng, N.~V. Prokof'ev, B.~V.
  Svistunov, and E.~Kozik, ``Extended crossover from a fermi liquid to a
  quasiantiferromagnet in the half-filled 2d hubbard model,'' {\em Phys. Rev.
  Lett.}, vol.~124, p.~017003, Jan 2020.

\bibitem{kim_cdet}
A.~J. Kim, F.~Šimkovic IV, and E.~Kozik, ``Spin and charge correlations across
  the metal-to-insulator crossover in the half-filled 2d hubbard model,'' {\em
  Phys. Rev. Lett.}, vol.~124, p.~117602, Mar 2020.

\bibitem{lenihan2020entropy}
C.~Lenihan, A.~J. Kim, F.~Šimkovic IV, E.~Kozik, {\em et~al.}, ``Entropy in
  the non-fermi-liquid regime of the doped $2 d $ hubbard model,'' {\em arXiv
  preprint arXiv:2001.09948}, 2020.

\bibitem{feldman}
F.~Feldman, H.~Kn\"orrer, M.~Salmhofer, and E.~Trubowitz, ``The temperature
  zero limit,'' {\em Journal of Statistical Physics}, vol.~94, pp.~113--157,
  1999.

\bibitem{rubtsov2005continuous}
A.~N. Rubtsov, V.~V. Savkin, and A.~I. Lichtenstein, ``Continuous-time quantum
  monte carlo method for fermions,'' {\em Physical Review B}, vol.~72, no.~3,
  p.~035122, 2005.

\bibitem{rossi2020renormalized}
R.~Rossi, F.~Šimkovic IV, and M.~Ferrero, ``Renormalized perturbation theory
  at large expansion orders,'' {\em arXiv preprint arXiv:2001.09133}, 2020.

\bibitem{bohm1951collectiveI}
D.~Bohm and D.~Pines, ``A collective description of electron interactions. i.
  magnetic interactions,'' {\em Physical Review}, vol.~82, no.~5, p.~625, 1951.

\bibitem{pines1952collectiveII}
D.~Pines and D.~Bohm, ``A collective description of electron interactions: Ii.
  collective vs individual particle aspects of the interactions,'' {\em
  Physical Review}, vol.~85, no.~2, p.~338, 1952.

\bibitem{bohm1953collectiveIII}
D.~Bohm and D.~Pines, ``A collective description of electron interactions: Iii.
  coulomb interactions in a degenerate electron gas,'' {\em Physical Review},
  vol.~92, no.~3, p.~609, 1953.

\bibitem{shifted_action}
R.~Rossi, F.~Werner, N.~Prokof'ev, and B.~Svistunov, ``Shifted-action expansion
  and applicability of dressed diagrammatic schemes,'' {\em Phys. Rev. B},
  vol.~93, p.~161102, Apr 2016.

\bibitem{kozik2015nonexistence}
E.~Kozik, M.~Ferrero, and A.~Georges, ``Nonexistence of the luttinger-ward
  functional and misleading convergence of skeleton diagrammatic series for
  hubbard-like models,'' {\em Physical review letters}, vol.~114, no.~15,
  p.~156402, 2015.

\bibitem{hubbard1963electron}
J.~Hubbard, ``Electron correlations in narrow energy bands,'' in {\em
  Proceedings of the royal society of london a: mathematical, physical and
  engineering sciences}, vol.~276, pp.~238--257, The Royal Society, 1963.

\bibitem{anderson1963theory}
P.~W. Anderson, ``Theory of magnetic exchange interactions: exchange in
  insulators and semiconductors,'' {\em Solid state physics}, vol.~14,
  pp.~99--214, 1963.

\bibitem{anderson1997theory}
P.~W. Anderson {\em et~al.}, {\em The theory of superconductivity in the
  high-Tc cuprate superconductors}, vol.~446.
\newblock Princeton University Press Princeton, NJ, 1997.

\bibitem{Note1}
We are grateful to T.~Ohgoe and F.~Werner for pointing out this fact to us.

\bibitem{griffin2006minors}
K.~Griffin and M.~J. Tsatsomeros, ``Principal minors, part i: A method for
  computing all the principal minors of a matrix,'' {\em Linear Algebra and its
  Applications}, vol.~419, no.~1, pp.~107 -- 124, 2006.

\bibitem{koivisto}
A.~Bj\"orklund, T.~Husfeldt, P.~Kaski, and M.~Koivisto, ``Fourier meets
  m\"obius: fast subset convolution,'' {\em Proceedings of the thirty-ninth
  annual ACM symposium on Theory of computing}, pp.~67--74, 2007.

\end{thebibliography}

\end{document}